\RequirePackage{fix-cm}
\documentclass[twocolumn,epjc3]{svjour3}  
\smartqed  
\RequirePackage{graphicx}
\usepackage{siunitx}
\usepackage{amssymb}
\usepackage{xcolor}
\journalname{Eur. Phys. J. C}
\usepackage[]{lineno}
\usepackage{hyperref}
\usepackage{ulem}
\begin{document}

\title{An Experiment for Electron-Hadron Scattering at the LHC
}



\long\def\comment#1{}
\author{K.~D.~J.~Andr\'e\thanksref{addr3,addr5}
        \and
        L.~Aperio Bella\thanksref{addr223}
        \and
        N.~Armesto\thanksref{e1,addr1}
        \and
        S.~A.~Bogacz\thanksref{addr4}
        \and
        D.~Britzger\thanksref{addr222}        
        \and
        O.~S.~Br\"{u}ning\thanksref{addr3}
        \and
        M.~D'Onofrio\thanksref{addr5}
        \and
        E.~G.~Ferreiro\thanksref{addr1} 
        \and
        O.~Fischer\thanksref{addr5}
        \and
        C.~Gwenlan\thanksref{addr224}
        \and
        B.~J.~Holzer\thanksref{addr3}
        \and
        M.~Klein\thanksref{addr5}
        \and
        U.~Klein\thanksref{addr5}
        \and
        F.~Kocak\thanksref{addr6}
        \and
        P.~Kostka\thanksref{addr5}
        \and
        M.~Kumar\thanksref{addr7} 
        \and
        B.~Mellado\thanksref{addr7,addr8}
        \and
        J.~G.~Milhano\thanksref{addr2,addr2222}
        \and
        P.~R.~Newman\thanksref{addr221}
        \and
        K.~Piotrzkowski\thanksref{addr1077}
        \and
        A.~Polini\thanksref{addr118}
        \and
        X.~Ruan\thanksref{addr7}
        \and
        S.~Russenschuk\thanksref{addr3}
        \and
        C.~Schwanenberger\thanksref{addr223}
        \and
        E.~Vilella-Figueras\thanksref{addr5}
        \and
        Y.~Yamazaki\thanksref{addr95}
}

\thankstext{e1}{e-mail: nestor.armesto@usc.es}


\institute{CERN, Esplanade des particules 1, 1211 Geneva 23, CH\label{addr3}
            \and
University of Liverpool, Oxford Street, UK-L69 7ZE Liverpool, United Kingdom \label{addr5}
            \and
            Deutsches Elektronen-Synchrotron (DESY), Notkestr. 85, 22769 Hamburg, Germany\label{addr223}
            \and
            Instituto Galego de F\'{\i}sica de Altas Enerx\'{\i}as IGFAE, Universidade de Santiago de Compostela, 15782 Santiago de Compostela, Galicia-Spain \label{addr1}
            \and
            JLab,  Newport News, Virginia, USA\label{addr4}
            \and
            Max-Planck-Institut f\"ur Physik, F\"ohringer Ring 6, 80805 M\"unchen, Germany \label{addr222}
            \and
            Department of Physics, The University of Oxford, Oxford, OX1 3PU, United Kingdom\label{addr224}
            \and
            Bursa Uludag University, Bursa, Turkey\label{addr6}
            \and
            School of Physics and Institute for Collider Particle Physics, University of the Witwatersrand, Johannesburg, Wits 2050, South Africa.\label{addr7}
            \and 
            iThemba LABS, National Research Foundation, PO Box 722, Somerset West 7129, South Africa. \label{addr8}
            \and 
			Instituto Superior T\'ecnico (IST), Universidade de Lisboa, Av. Rovisco Pais 1, 1049-001, Lisboa, Portugal\label{addr2}
			\and
            LIP, Av. Prof. Gama Pinto, 2, P-1649-003 Lisboa , Portugal \label{addr2222}
			\and
            School of Physics and Astronomy, University of Birmingham, UK \label{addr221}
            \and
            Universit\'e Catholique de Louvain, Centre for Cosmology,
Particle Physics and Phenomenology,
1348 Louvain-la-Neuve, Belgium\label{addr1077}
            \and
            Istituto Nazionale di Fisica Nucleare (INFN), Sezione di Bologna, Bologna, Italy\label{addr118}
           \and
            Graduate School of Science, Kobe University, Rokkodai-cho 1-1, Nada, 657-8501 Kobe, Japan\label{addr95}
 }   
            


\date{Received: date / Accepted: date}

\maketitle

\begin{abstract}
Novel considerations are presented 
on the physics, apparatus and accelerator designs for a future, luminous, energy frontier electron-hadron ($eh$) scattering experiment at the LHC in the thirties for which 
key physics topics  and 
their relation to the hadron-hadron HL-LHC physics programme are discussed.
Demands are derived set by these physics topics on the design of the LHeC detector,
a corresponding update of which is described. Optimisations on the accelerator design, especially the interaction region (IR),  are presented. 
Initial accelerator considerations 
indicate that a common IR is possible to be built which alternately 
could serve $eh$ and $hh$ collisions while other experiments would stay on 
$hh$ in either condition. A forward-backward symmetrised option of the LHeC detector
is sketched which would permit extending the LHeC physics programme to also include
aspects of hadron-hadron physics.
The vision of a joint $eh$ and $hh$ physics experiment is shown to open new prospects
for solving fundamental problems of high energy heavy-ion physics including 
the partonic structure of nuclei and the emergence of hydrodynamics in quantum field theory while the
genuine TeV scale DIS physics is of unprecedented rank.

\end{abstract}

\section{Introduction}
\label{intro}

The Standard Model (SM) of particle physics is based on a non-Abelian gauge theory with a symmetry group SU(2)$_L\times$U(1)$\times$SU$_c$(3). The SM has and continues to enjoy great success in describing a wide span of phenomena emerging from interactions of particles at a range of energies that is accessible experimentally. That said, the SM is not a satisfactory theory of fundamental interactions nor does it explain a number of phenomena in nature. It is of paramount importance to the field of particle physics to establish
whether and how the SM breaks down in laboratory conditions.
This is expected to be achieved by pushing the boundaries of energy and precision frontiers, and through various sensitive experiments at low energy. Theory is currently less predictive than ever after the birth of the SM, such that experimentation based on novel designs acquires a particular eminence for the decades ahead.  

Deep inelastic scattering (DIS) of electrons and photoproduction off high energy protons and ions
with high instantaneous luminosity offers a unique opportunity to enhance the precision frontier in particle physics, for which examples are provided in this paper. The intense, unique hadron beams of the LHC represent a salient opportunity to 
create a new laboratory for particle physics and energy frontier DIS,
the Large Hadron electron Collider (LHeC),
at affordable cost: a larger than TeV centre of mass system (cms) energy new collider is in sight by adding an energy recovery linac to the LHC, in possibly staged 
phases. The present paper is mainly devoted to an update of the detector, describing relevant  physics, apparatus and accelerator design considerations and new results. The LHeC would be the fifth large collider experiment at the LHC facility, sustaining its future and exploiting
the hitherto biggest investment in particle physics.

A first comprehensive design concept for the LHeC was published in 2012~\cite{AbelleiraFernandez:2012cc}, just weeks prior to the Higgs boson ($H$) discovery and incorporating the findings of a review pursued by twenty experts in experimental, theoretical or accelerator physics.  
Following nearly ten years of LHC operation and analysis, incorporating technology
progress, accounting for the advent of experimental Higgs physics and relying 
on the brilliant LHC performance, a further detailed report appeared recently, 
written again by representatives of more than a hundred institutions~\cite{Agostini:2020fmq}.
That paper presented the energy recovery, linac ring electron-hadron collider  as the default configuration with luminosity parameters an order of magnitude enhanced compared to before. 
It suggested to downscope the electron beam energy from originally 60 to 50\,GeV in an
attempt to economise both investments and efforts as the racetrack electron accelerator circumference then became comparable to that of the SPS\footnote{Pending on the finally achievable operational
gradients in the SRF linac, an upgrade to 60\,GeV might still be feasible at a second phase of the LHeC operation.}. This concept, its
main ingredients, had been submitted to the European Strategy Update and published a year before the extended paper~\cite{Bruning:2019scy}.

The LHeC has been maintained as an option and complement of HL-LHC in strategic considerations of the future, especially the forthcoming European road-maps on detector
and accelerator research and development. Its main role, naturally, is that 
of complementing the TeV scale exploration with the LHC and a possible future $e^+e^-$ collider, much like HERA was coupled to the Tevatron and LEP before.

The present paper summarises key physics topics, on hadron structure,
parton dynamics, Higgs, top and BSM physics and it enlists the detector requirements
derived from each of these topics leading to an update of the LHeC apparatus concept which is here
detailed.

For long, the LHeC development followed the common understanding that with the long shutdown (LS) 4, in the early thirties, the operation of the LHC as a heavy-ion (HI) collider would be terminated in order to maximise $pp$ luminosity, which would have enabled using the Interaction Point (IP) 2 for a new experiment.
Meanwhile one considers operating LHC with heavy ions further hence. New considerations have appeared for a follow-up heavy-ion experiment at IP2, 
configured to study soft heavy-ion interactions~\cite{Adamova:2019vkf}, while
heavy-ion physics is newly discussed at LHCb and ATLAS and CMS continue
their HI programmes too.

In order to conceptually resolve a possible clash of the LHeC and ALICE3 plans in a productive manner,
 it had been suggested by the International Advisory Committee on the LHeC, see Appendix  in~\cite{Agostini:2020fmq} and ~\cite{ECFANewsL5}, to investigate whether an adjusted 
 LHeC detector, or a commonly designed apparatus~\footnote{While this paper, including the concept of a combined $eh$ and $hh$ detector at IP2, was in the peer-review process, the concept of the ALICE3 detector is evolving, possibly extending the initial tracking device concept~\cite{Adamova:2019vkf}  to also 
include an electromagnetic calorimeter, a higher field solenoid and a muon spectrometer following an absorber~\cite{newA3}.}, and the interaction region (IR) could be configured to register both $ep/eA$ and, alternately, $AA$ scattering events. 
 This paper comprises an initial study of the accelerator and IR modifications which
 suggest that a common IR may indeed be feasible, with features that are
 further to be evaluated.
   In this context, a symmetrised first version of the LHeC detector is here presented
 together with a discussion of a future joint $eh$ and $hh$ heavy ion physics programme.

This article is devoted to experimenting at the LHeC with the option of alternately registering electron-hadron ($p,~A$) and hadron-hadron interactions in a common detector. 
Section~\ref{physics} presents key physics topics in DIS scattering with emphasis
on constraints on the detector layout and performance.  Section~\ref{detector} presents
an updated LHeC detector layout, with an enlarged tracker radius and choosing
liquid argon technology for the electromagnetic calorimeter. A detailed evaluation of
a variety of aspects of the accelerator design, especially the IR and its possible
configuration to alternately serve $eh$ and $hh$ interactions, is given in 
Section~\ref{machine}. Section~\ref{jointehhh} is dedicated to a comprehensive description of
the heavy-ion physics potential for a joint $eh/hh$ experiment and a brief presentation 
of a symmetrised version of the LHeC detector to enable its use for $hh$ physics.
The paper concludes with a summary in Section~\ref{summary}.

\section{Physics}
\label{physics}

The physics programme at the LHC and the DIS programme at the LHeC are both extremely rich and stand on their own. However,
they also have much in common: with the necessity of understanding hadron structure and parton dynamics
for searches for new physics and precision measurements at the LHC, with 
the
opportunity to explore the Higgs mechanism at per cent level,  novel top quark physics, precision SM measurements and further, in the search for new physics and in the understanding of nuclear parton structure and the phenomenon of the Quark Gluon Plasma and heavy ion physics in general.
With a view on the resulting detector constraints
and for illustrating the exciting physics programme that the LHeC entails, we have chosen these key topics for a brief
description of the potential of the ``Experiment for
Electron-Hadron Scattering at the LHC" we here describe. Some special emphasis is given
to heavy ion physics in view of the idea, mentioned above, of possibly realising this experiment in a configuration that may jointly be used by DIS oriented and  more heavy ion interested communities.
Similar illustrations of the physics potential and experimental requirements could be provided for the physics at small Bjorken $x$, for diffraction, photo-production physics
and other areas,
see ref.~\cite{Agostini:2020fmq}.
One should be aware that the opening of an unexplored kinematic range, accessed with so high luminosity, may lead to 
surprises and should not pretend to be able to predict everything. This also regards, technically, the development of analysis tools, for which the past decade on LHC physics brought many examples of results exceeding 
in their depth and precision the initial expectations by far. Finally, new theoretical insight or surprises from other particle physics experiments, may indeed shift the focus.

\subsection{Partons and Proton Structure}
\label{sec:partons}
One may distinguish four phases, including the LHeC,
of the experimental development of the physics of parton structure of the proton which was opened with the 
SLAC-MIT lepton-hadron scattering experiment at Stanford in 1968: fixed target experiments, HERA, $pp$ Drell-Yan scattering and the LHeC. The role of a next, luminous energy frontier $ep$ scattering experiment becomes obvious when
one revisits the past and realises the unique potential
of the LHeC, recently presented in much detail~\cite{Agostini:2020fmq}.

Partons, quarks and gluons, are confined inside the proton; 
still a major puzzle for modern physics, they cannot be observed directly.  Quarks
of any type $q$ (and $\bar{q}$) have a certain probability of carrying a fraction $x$ of the proton's momentum,  
described by a momentum density function $xq(x)$, called a parton distribution function (PDF).
The characteristics of the proton are  given by the valence quark content of two up 
and one down quarks. 
The relative distribution of the proton's momentum among the quarks varies with $x$.  
It changes as we resolve the proton more deeply in lepton-hadron deep inelastic scattering,
i.e. through a virtual photon or a $Z$ or $W^{\pm}$ boson, of virtuality $Q^2$, 
interacting with a quark. 
The strong interaction between quarks is mediated by
gluons, discovered in 3-jet events in $e^+e^-$, which carry a half of the proton's momentum. 
The quark-gluon interactions are described within 
QCD, with a coupling constant $\alpha_s(Q^2)$. With rising $Q^2$, the coupling decreases 
logarithmically such that asymptotically quarks become free and the strong interaction
at the parton level can be described as a perturbation theory. These and further
fundamental properties have been established\footnote{Despite their phenomenological success there continue to exist certain doubts about the whole parton picture based on principles for the structure of nature going back to Newton~\cite{Yock:2020kli} with testable hypotheses at the LHeC, which therefore has been termed the ``Newtonian Telescope of CERN".} in a first era of PDF physics enabled
by a series of neutrino, electron and muon scattering experiments on stationary hadronic targets.

HERA was the first $ep$ collider. It extended the
kinematic range of DIS experiments, given as $s=Q^2_{max}=4E_eE_p$, by two orders of magnitude
but fell short against those by another about two orders of magnitude in luminosity.
 Its contributions to the understanding of parton structure and dynamics, nevertheless, can 
 not be underestimated. Of special importance
 has been, firstly, the extension of 
 DIS into the very high $Q^2$ region with a) the validation of the linear DGLAP 
 evolution law to $Q^2$ values beyond the weak boson 
masses, $\gtrsim 10^4$\,GeV$^2$, and b) the simultaneous
use within one experiment of the charged (CC) and neutral current (NC) weak interaction, besides the 
electromagnetic NC photon exchange, to determine PDFs,
including first determinations of the charm and bottom
quark densities through impact parameter measurements.
Since the accessible $x$ range towards small
$x$ is extended $\propto 1/s$, HERA was able to resolve, in addition,
the gluon, sea and valence quark behaviour at small $x$. It
established the dominance of the gluon density $xg$ at small $x$ but could not
convincingly answer the question of whether non-linear gluon-gluon 
interactions occur, which would damp the rise of $xg$ towards small $x$ and lead outside 
the validity range of the DGLAP equations. The HERA NC and CC collider data did 
permit a first and far reaching set of PDFs to be determined, 
without using additional data, with their canonical 
uncertainties understood at the $\Delta \chi^2 =1$ level~\cite{H1:2015pak}, 
and they are the inevitable
part of any modern PDF determination.

Following HERA, with the advent of the LHC and its Drell-Yan
measurements, the art of extracting PDFs from so-called global data has become an active field of particle physics,
to test QCD and to understand LHC measurements using maximum suitable data, and novel analysis and mathematical
methods. Such analyses carry a number of severe theoretical
and practical difficulties which, despite impressive
successes by the various PDF analysis groups, lead to a principally unsatisfactory situation due to the 
nature of hadron-hadron scattering with respect to DIS. This comprises effects of initial state quark radiation (leading to resummation), of
hadronisation and reconstruction arbitrariness in jet data, the incompatibility of many data sets leading to the rather ad-hoc inflation of uncertainty bands or even exclusion of the most precise data, such as the ATLAS inclusive W,Z data from CT18~\cite{Hou:2019efy}, for example.  A reflection of these effects is the observed difference between PDF sets of different groups which is often larger than the claimed precision of fits.  
A conceptual difficulty is the uncertainty at high mass, corresponding to large $x \gtrsim 0.5$, where the occurrence of new physics is possible,  such that the LHC data should be excluded from PDF fits. The current status of the determination of $\alpha_s$ to about 2\,\% uncertainty 
limits the PDF determination, and the precision of predictions such as the $gg \to H$ production cross section, being $\propto \alpha_s^2$. Simulation studies on future PDF determinations from the LHC assume that the data compatibility problems may disappear, while the principal problems will in fact remain. 

A precision physics era at the HL-LHC will be maximally precise if it was accompanied by the LHeC PDF 
programme, the possible fourth phase of PDF physics. As described in detail in~\cite{Agostini:2020fmq}: i) the 
increased energy will make the CC DIS data for the first time a useful base extending over 4 orders of magnitude in $x$ and $Q^2$; ii) all PDFs, $xq(x,Q^2)$ and $xg(x,Q^2)$,
\footnote{PDFs depend on a factorisation scale $\mu_F^2$ which in DIS is taken to be equal to $Q^2$. Also note that we make no distinction between Bjorken $x$ as the kinematic variable in DIS and $x$ as the momentum fraction of the proton carried by the probed parton as they coincide in the parton model.}, 
can be determined in a single DIS experiment over many orders of magnitude, with
$q=u_v,~d_v,~u,~\bar{u},~d,~\bar{d},~s,~c,~b$ and also $t$;
iii) the kinematic range, unlike at HERA or lower energy
fixed target or $ep$ collider experiments, extends to such low values of $x$ in the DIS region that one will be able to settle the question of non-linear gluon-gluon interactions, etc. An unprecedented precision on these distributions is in reach, as has been simulated more than once but conclusively in~\cite{Agostini:2020fmq}, including per mille accuracy of $\alpha_s$. This will comprehensively test pQCD and the underlying parton dynamics view at the highest level; will enable new physics, possibly occurring in the high mass tails from interference contact interaction effects, to be discovered at LHC; lead to possible discoveries in QCD such as the breaking of factorisation not only in diffraction; and enable precision electroweak and Higgs physics at the joint
$ep/pp$ LHC facility to a stunning level of precision. 

\begin{figure}[htb]
\begin{center}
    \includegraphics[width=0.23\textwidth]{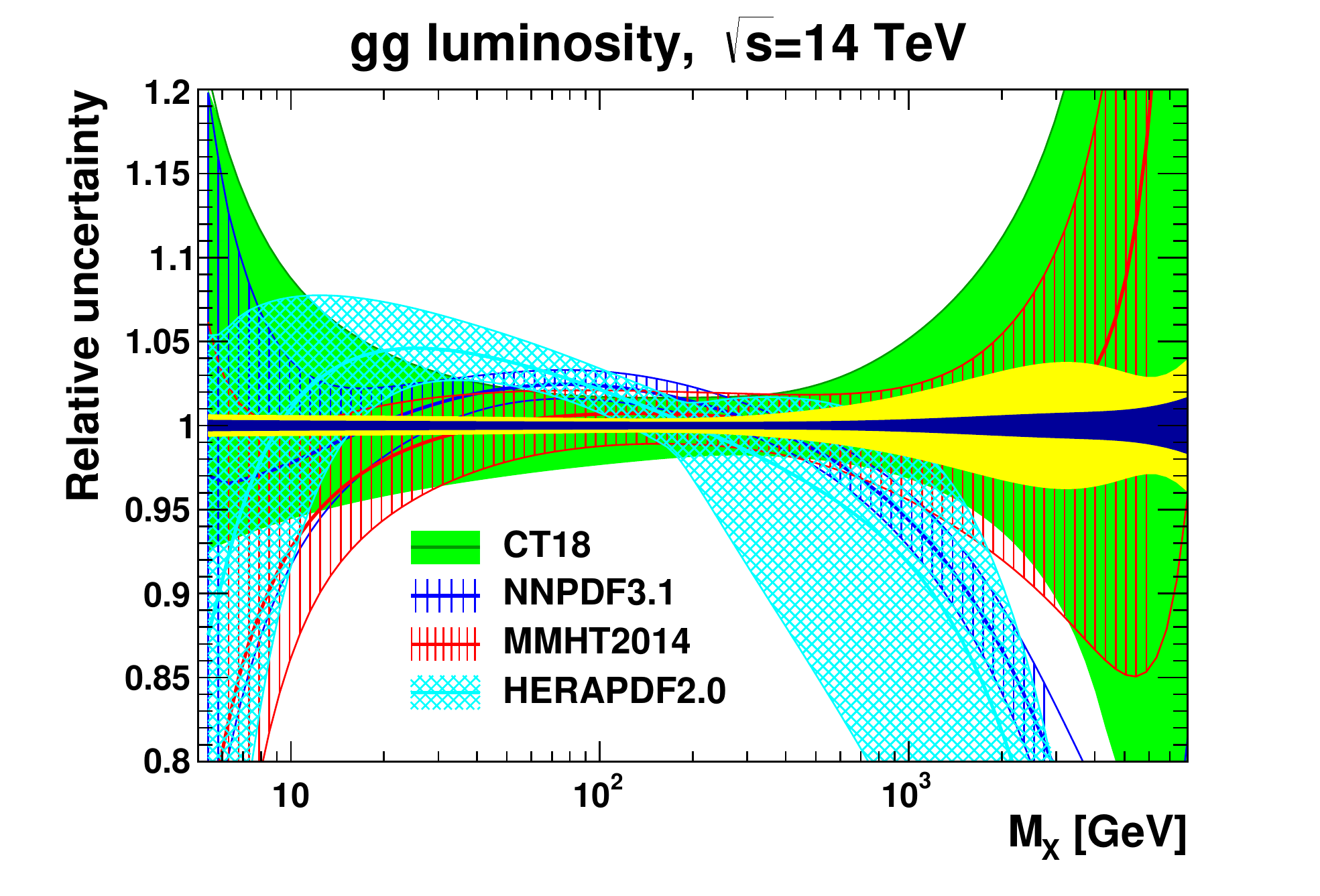}
    \includegraphics[width=0.23\textwidth]{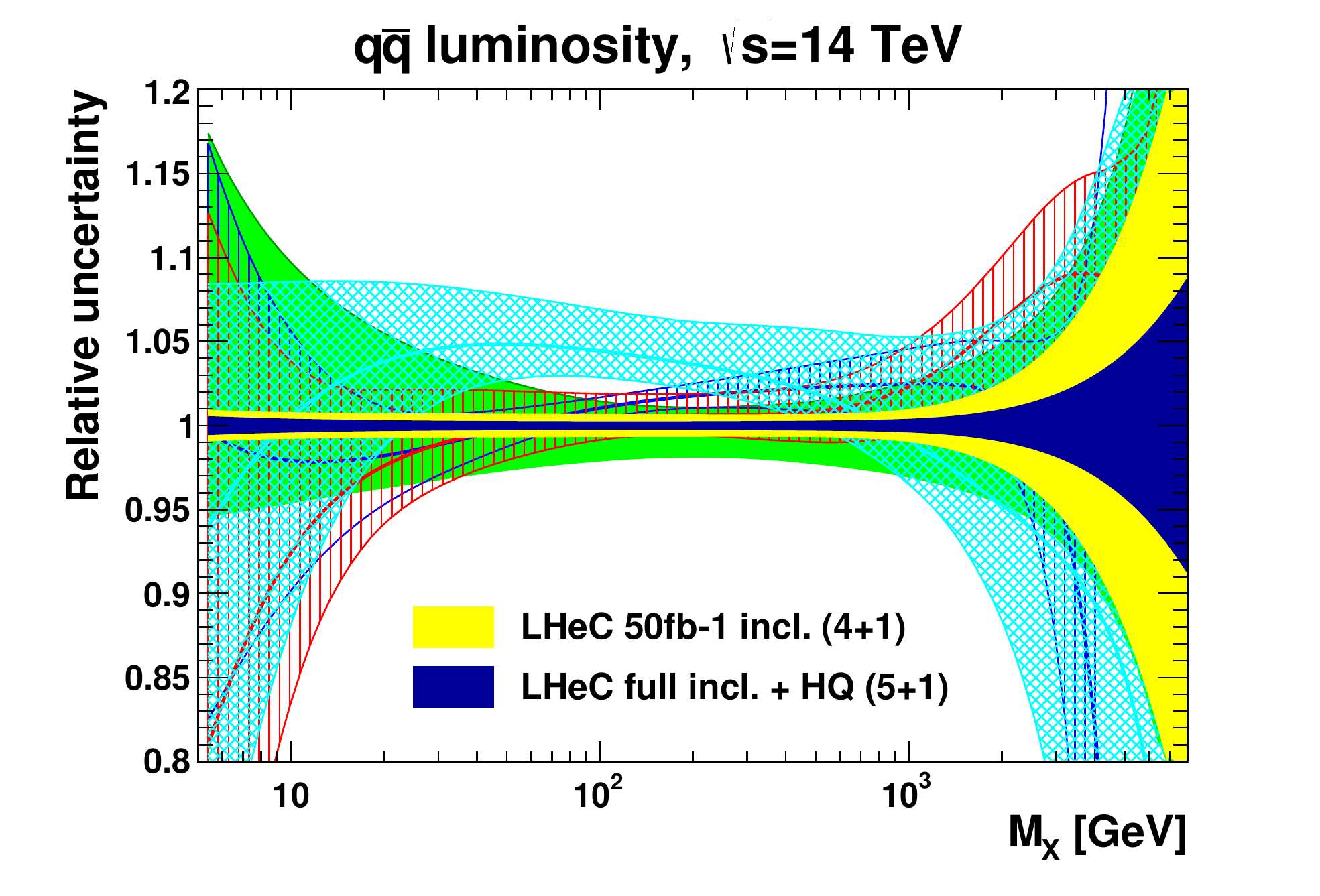} \\
    \includegraphics[width=0.23\textwidth]{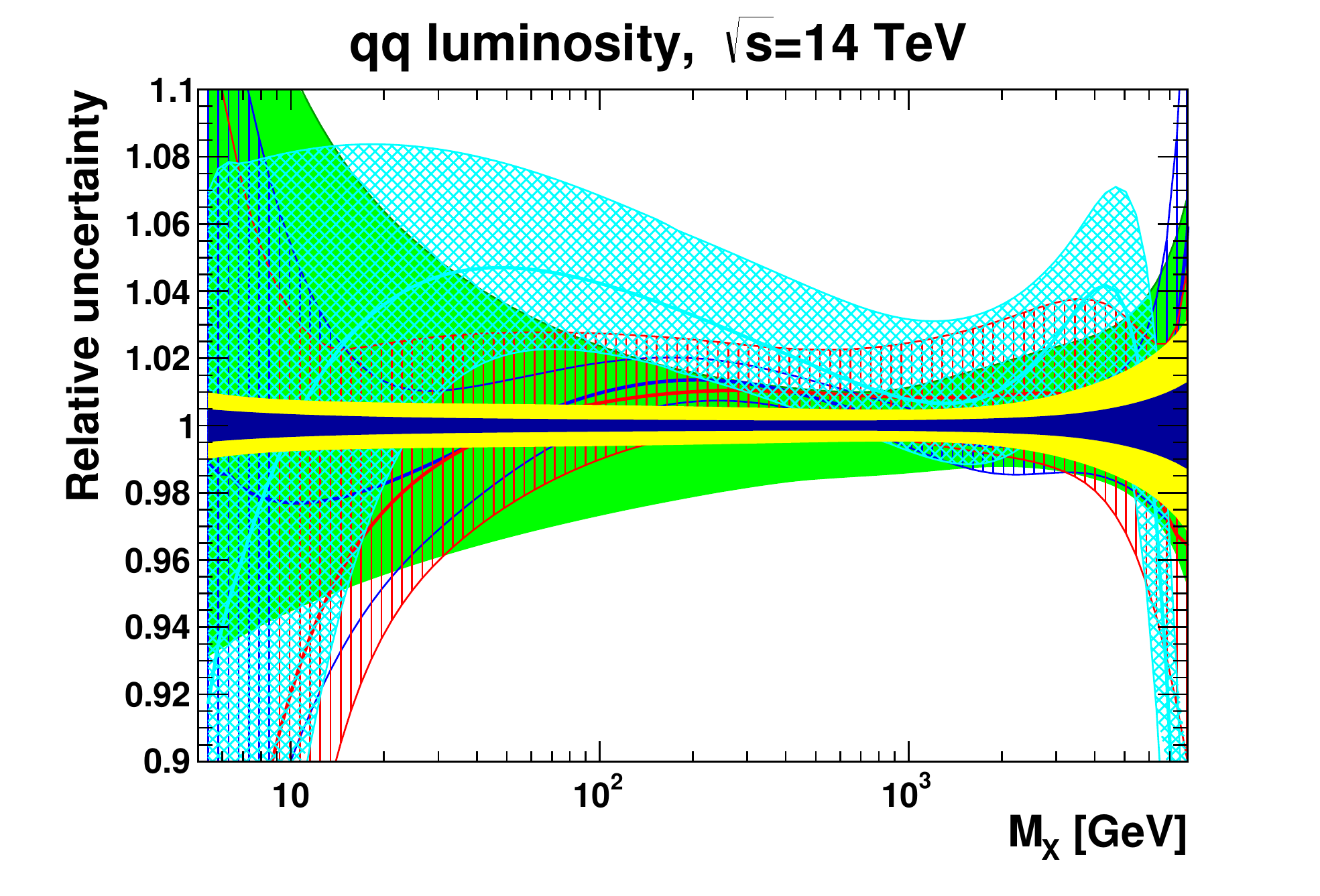}
   \includegraphics[width=0.23\textwidth]{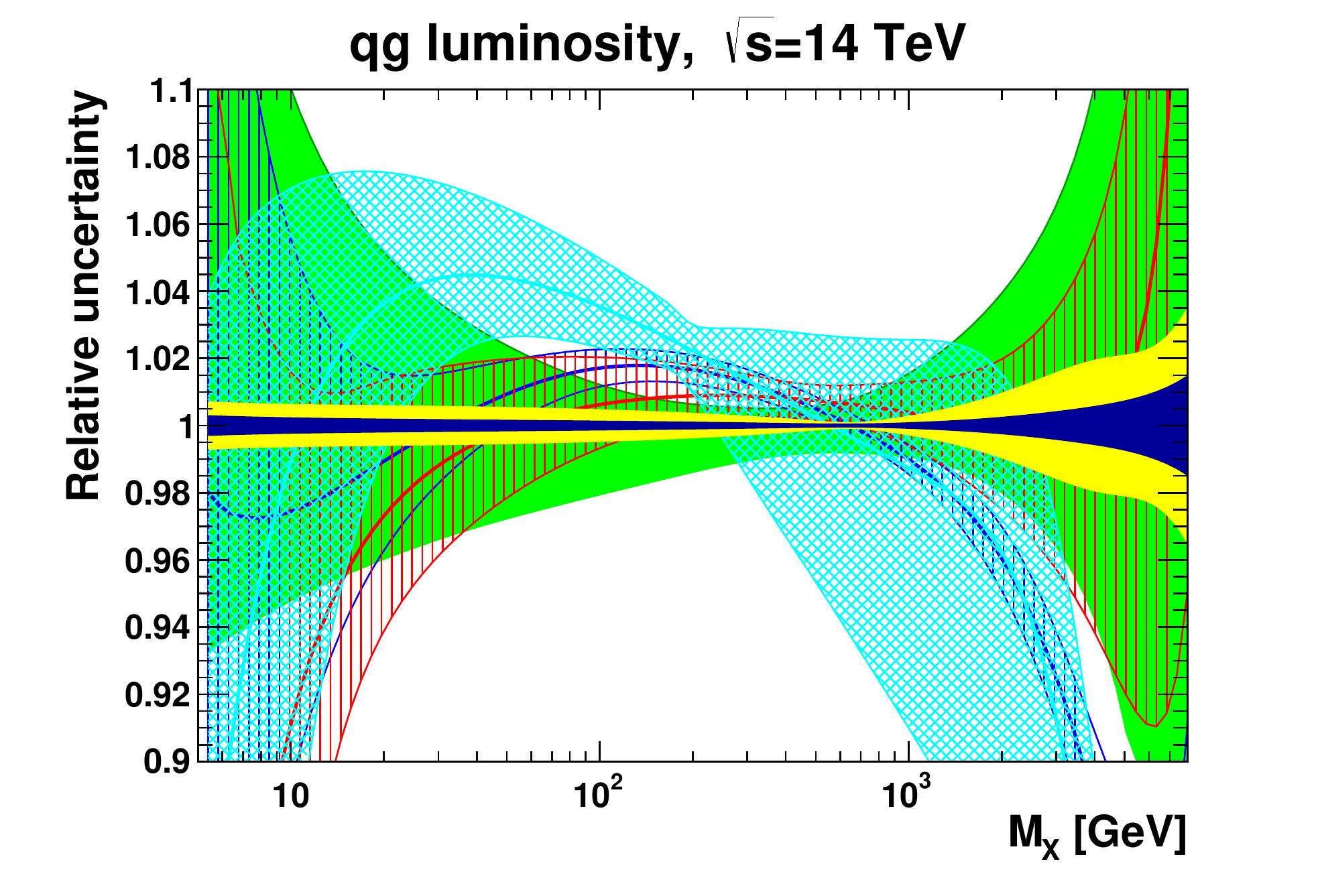}
  \end{center}
\caption{Expected precision for the determination of
parton-parton luminosities as function of 
$M_X$ in Drell-Yan scattering at the 14 TeV LHC. Light blue: HERA, yellow: initial LHeC run, dark blue: full LHeC data set, overlayed with three, recent global fit results. For more information see~\cite{Agostini:2020fmq}.}
\label{fig:pplumi}       
\end{figure}

Such an ambitious programme, including precision measurements of the strange, charm and bottom quark distributions and of the longitudinal structure function
$F_L(x,Q^2)$, sets important constraints for the experiment here presented: i) it is very desirable that such data exist while HL-LHC operates. Therefore a dedicated study~\cite{Agostini:2020fmq} has been made of the LHeC PDF prospects for an initial data set of $50$\,fb$^{-1}$, see Fig\,\ref{fig:pplumi}.
Such a luminosity is a factor of 100 larger than that which H1 and ZEUS collected in their 
15 year lifetime, while being expected in the first LHeC running period~\cite{Bordry:2018gri}; ii) the detector acceptance should extend maximally to small hadron final state angles to cover larger $x$ and to low electron scattering angles to cover low $Q \sim 1$\,GeV$^2$, even when one can extend the region of acceptance considerably with
lower beam energy runs; iii) hermiticity of the apparatus is required to apply an $E-p_z$ balance criterion which diminishes the radiative corrections substantially; iv) cross calibration of the hadronic and electromagnetic calorimeter as well as polar angle measurements should ensure a below per cent level accuracy of the energy scales keeping the experimental scale uncertainties small; v) high resolution
hadron
energy
measurements are important for controlling hadronic backgrounds and reconstructing missing energy; vi) heavy flavour reconstruction requires impact parameter resolutions of order 10\,$\mu$m resulting from novel tracking technology and the small beam size of about 
$7$\,$\mu$m transversally, twenty times better than at HERA; vii) the large photo-production background shall be tagged,
for its own physics study and for subtracting it in DIS measurements. A major demand in $ep$ scattering is the
control of halo and synchrotron radiation backgrounds through a carefully designed interaction region, see
Sect.\,\ref{sec:ir}.

\subsection{Higgs Boson Physics}
\label{sec:Higgs}

The Higgs boson is of fundamental importance as it is related
to the spontaneous breaking of a locally symmetric gauge theory, to a mechanism 
in which the
intermediate vector bosons are explained to be massive while 
the photon is kept massless. Fermions acquire a mass through the Higgs field. This spectacular theoretical prediction was confirmed
in 2012 at the LHC, which boosted Higgs physics to the top
of particle physics investigations. The nearest task is to verify 
the theory in more detail, especially to reconstruct the complete
decay spectrum of the Higgs boson, besides searching for possible
extensions of the Higgs mechanism and relations of the Higgs boson to exotic particles. 

Prior to the discovery of the Higgs boson, initial work on the Higgs physics potential 
in DIS was centered around
the dominant decay channel into $b \bar{b}$~\cite{Han:2009pe}, and a first
comprehensive study was published with the LHeC design report in 
2012~\cite{AbelleiraFernandez:2012cc}. It had been realised that the theoretical
understanding of Higgs production in $ep$ was particularly suited for
precision measurements with small QED and QCD corrections~\cite{Blumlein:1992eh,Jager:2010zm}.  

\begin{figure}[th]
\centering
\includegraphics[width=0.7\textwidth]{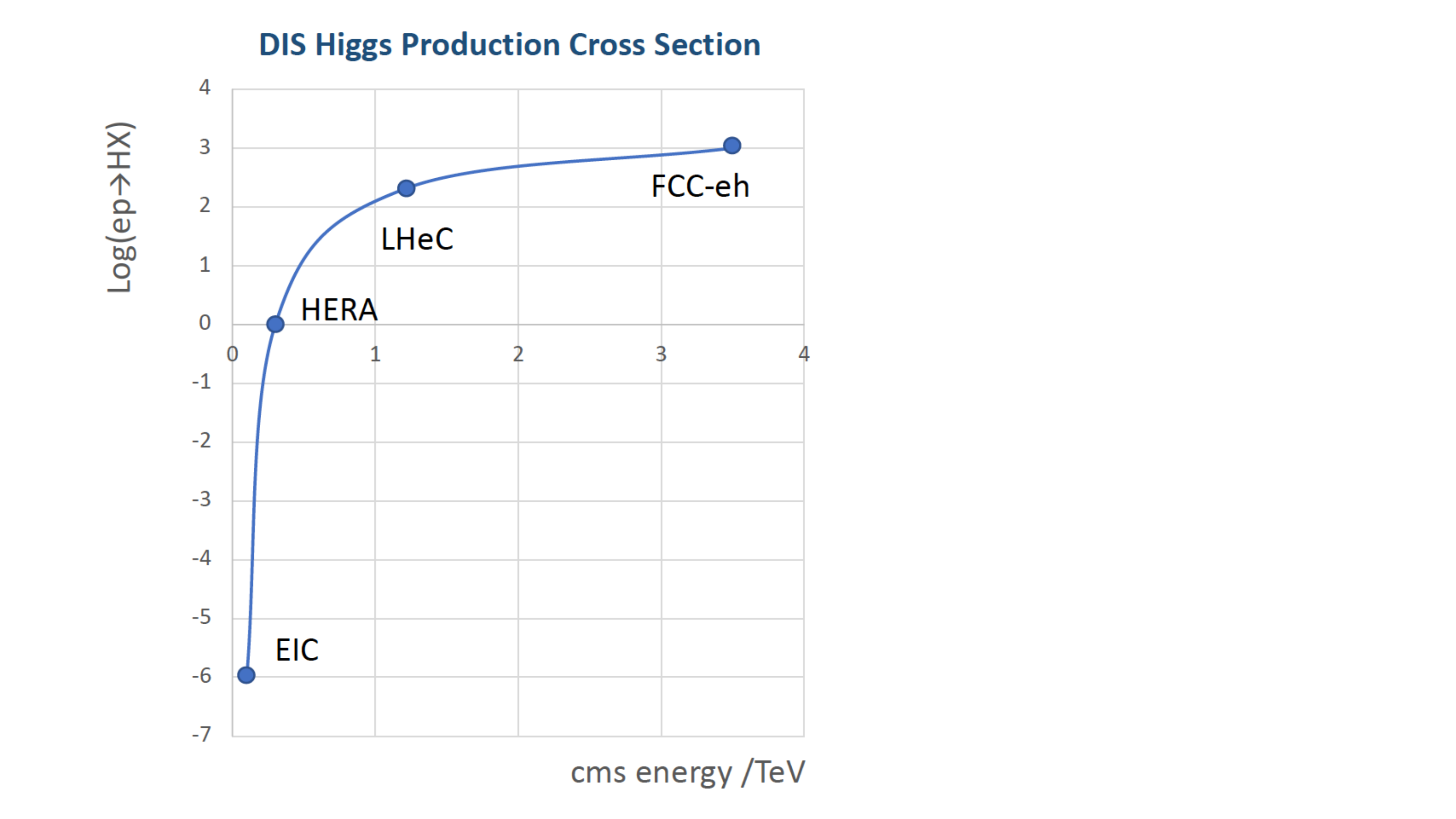}
\caption{
Inclusive Higgs cross section in charged current $e^-p$ DIS, plotted
as $\log{(\sigma /fb)}$, as a function of the cms $ep$ scattering
energy, $\sqrt{s} = 2 \sqrt{E_eE_p}$ in TeV. 
}
\label{fig:Hsec}
\end{figure}

In deep inelastic scattering, the Higgs boson is produced predominantly
in charged current reactions, thro\-ugh the $t$-channel reaction $WW \to H$, with the neutral current cross section being smaller but still measurable. The production cross section depends strongly on the cms energy as is shown in Fig.\,\ref{fig:Hsec}.
The cross section
at the LHeC is about 200\,fb, which is close to
 the $e^+e^- \to H Z$ cross section in 
electron-positron scattering at 250\,GeV energy. 
With thousand times its luminosity, HERA would have had a chance to
observe the Higgs boson in DIS
while
that is beyond the reach of lower energy $ep$ colliders such as the EIC. 
At the FCC-eh the SM H boson is calculated to have  a pb cross section which implies it has an outstanding potential for Higgs physics as has been demonstrated in~\cite{FCC:2018xdg}.

A comprehensive analysis of  SM Higgs boson physics at the LHeC and the FCC-eh has recently been presented in~\cite{Agostini:2020fmq}.
The LHeC is sensitive to the six most frequent decay channels,
$b \bar{b},~W^+W^-,~gg,~\tau \tau,~c\bar{c}$ and $ZZ$, which represent about $99.6$\,\% of the total SM Higgs boson decay width. Owing to the high precision with which the couplings to the $b$-quark and the $W$ and $Z$ bosons can be measured, the latter cleanly
in $WW \to H \to WW$ in CC and $ZZ \to H \to ZZ$ in NC, 
one finds that the total of the (six) SM decays can be reconstructed
at per cent level accuracy. 

\begin{figure}[th]
\centering
\includegraphics[width=0.58\textwidth]{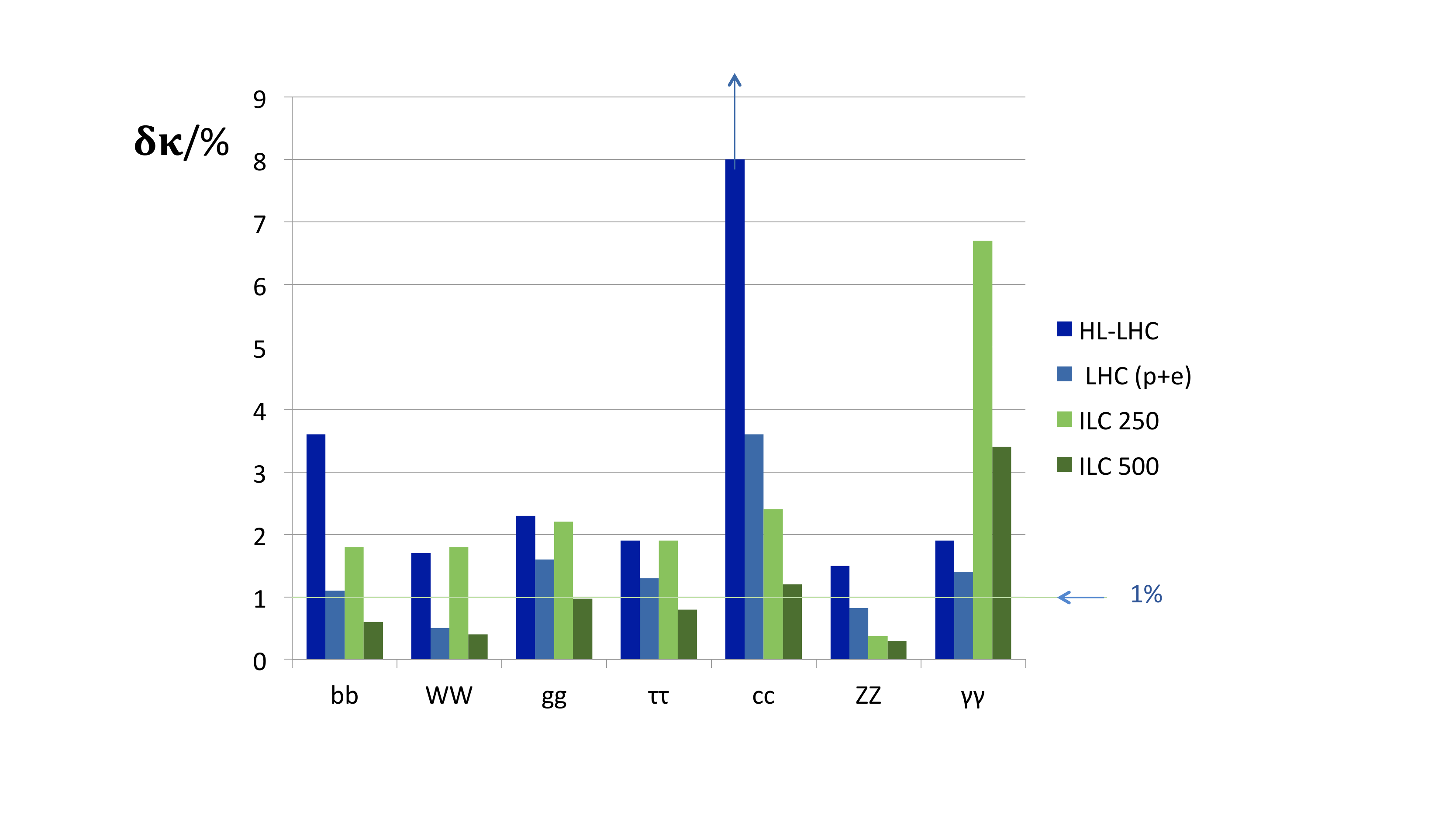}
\vspace{-0.5cm}
\caption{Results of prospect evaluations of the determination of Higgs couplings in the SM kappa
framework for HL-LHC (dark blue), LHC with LHeC combined (p+e, light blue), ILC 250 (light
green) and ILC-500 (dark green)~\cite{Fujii:2017vwa}.
}
\label{fig:hall}
\end{figure}

A dedicated analysis has also been presented in~\cite{Agostini:2020fmq} of the potential which the LHC facility at large had for precision
Higgs physics if one was eventually able to combine the HL-LHC results with the ones from LHeC. This analysis is illustrated in Fig.\,\ref{fig:hall} 
as a comparison of the LHC facility, singly $pp$ and $ep~\&~pp$, with the 
International Linear Collider (ILC) in its initial and an upgraded configuration. 
Two observations may be emphasised: 
i) the addition of the LHeC to the LHC improves the proton-proton result very significantly for the dominant decay channels, $bb$ and $WW$~\footnote{This
underlines the observation that the $ep$ configuration has a unique sensitivity 
to the $HWW$ vertex, which may lead to the observation of anomalous effects when
measured with high precision~\cite{Biswal:2012mp}.},
also for $ZZ$. The improvements on the gluon and tau channels arise from measurements which are roughly as precise in $ep$ as in $pp$. A striking result
is the expectation to measure the charm coupling accurately in $ep$, to $\delta \kappa_c \simeq 4$\,\%, which is considered to be unaccessible in $pp$ scattering
for large combinatorial background reasons; ii) the combination of LHC and LHeC promises to deliver results some which are better than those at the ILC, and vice versa too. One observes an intriguing complementarity between $ep$ and $e^+e^-$. The dominant production mechanisms are CC $ep$ scattering and $Z$ Higgsstrahlung, which explains why $\kappa_W$ is 
estimated to be most precise in $DIS$
and $\kappa_Z$ in $e^+e^-$. Similar observations have been made in the comparison of the FCC-eh and FCC-ee prospects at an even higher level of precision~\cite{FCC:2018xdg}.

Such comparisons have to be taken with care, they are the end product of advanced but prospect analyses, with many inherent accelerator and experiment assumptions. One yet may conclude that the one percent accuracy level for SM Higgs physics is indeed in reach for the hadron ($pp$ and $ep$) colliders, LHC and LHeC taken together, not worse than a next $e^+e^-$ collider
with a $10^{34}$ cm$^{-2}$s$^{-1}$ luminosity horizon, having the important additional feature to provide a model independent Higgs width measurement.

It has been emphasised~\cite{jorgeams} that a triangle of precise $pp$, $ep$ and $e^+e^-$ measurements is essential for resolving inherent correlations in the 
Higgs sector. i.e. there is more involved than a competition for highest luminosity prospects~\footnote{This is an example for LHC and ILC, there are more electron-positron collider options under discussion.
A comparative analysis of Higgs boson measurements at future colliders has been presented in~\cite{deBlas:2019rxi}. The field, i.e. theory, HL-LHC prospects, $e^+e^-$ collider prospects, see for example the ERL version 
of the ILC~\cite{1981219}, and other aspects, such as the EFT view on these measurements, is very dynamic and cannot be comprehensively described here.}.

As discussed below, the LHeC is a single top facility. The cross section for the production of the Higgs boson in association with a single top is sufficiently large for measurements to be affected. In the SM, the production of the Higgs production in association with a single top is heavily suppressed due to negative interference. As such, it is very difficult to access this production mechanism at the LHC. 

The LHeC provides a unique opportunity to study the CP structure of the Higgs boson Yukawa coupling \cite{Coleppa:2017rgb}. One can introduce CP-phase $\zeta_t$ of the $tth$ coupling, where $\zeta_t=0$ corresponds to the SM. As a result of the strong enhancement of  $pe^{-}\rightarrow \overline{t}h\nu_e$ for $\zeta_t>0$, strong limits can be set on deviations from the SM.

Assuming the Yukawa coupling to have the same structure as in the SM, the coupling size could be measured at the LHC with an accuracy of 17\% with 1\,ab$^{-1}$ of integrated luminosity~\cite{Coleppa:2017rgb}.  The use of multivariate techniques and additional channels, not employed so far, will further improve the precision of the measurement.  

For the LHeC accelerator, especially the luminosity targets, and the detector, 
the striking Higgs potential has had a strong influence on their designs. 
The following specific demands have been derived in shaping the LHeC detector 
design: i) b tagging is essential down to small angles corresponding to a pseudorapidity $\eta$ of up to about 3; ii) charm tagging requires an impact parameter resolution of $10~\mu$m or better, which raises interest in an instrumented beam pipe; iii) the hadronic calorimeter resolution should be as good as is feasible for reducing 
background; iv) efficient photo-production tagging of backward low $Q^2$ processes
is required for background rejection and normalisation; v) scattered electrons shall be identified up to $\eta \simeq 4$, similarly for muons which may emerge from abundant, exotic $H \to \mu \mu$ decays and vi) the reconstruction of missing energy poses
a desirable $\eta$ reach of up to 5. These considerations have been taken into account for the detector design. 

The reconstruction, especially of rarer channels, sets a goal of ${\cal O}(1)$\,ab$^{-1}$ of integrated luminosity, which is a desirable target too for BSM and very high $x$ and high $Q^2$ DIS physics. The Higgs potential of the LHeC and FCC-eh is striking and worth every attention already in the design phase, setting, for example, an about $20$\,mA current goal for the energy recovery facility PERLE~\cite{Angal-Kalinin:2017iup} which is being built at IJCLab Orsay.

\subsection{Top Quark Physics}
\label{sec:top}

Electron-proton colliders at high energy are ideal to study the electroweak interactions of the top quark. The LHeC is an outstanding single top facility in its own right. The charged current cross section is 1.9\,pb, compared to 0.05\,pb of the photo-production of $t\overline{t}$. This provides an opportunity to measure the $Wtb$ coupling with high precision and to search for anomalous contributions in the $Wtb$ vertex~\cite{Dutta:2013mva}. With 100\,fb$^{-1}$ of integrated $ep$ luminosity relative errors of order of 1\% can be achieved in the measurement of the $Wtb$ coupling. The Next-to-Leading Order corrections to the total and fiducial cross-sections are known~\cite{Gao:2021plf} and do not significantly affect the ability of the LHeC to achieve precision. These may reduce the expected fiducial cross-section of single top production by 14\%, while providing stability against scale variations. By contrast, measurements of single top production at the LHC are hampered by the large $t\overline{t}$ production cross-section. This is a further epitome of the complementarity of the LHeC and the LHC.

Given the high level of precision characteristic to the LHeC, other elements of the CKM matrix are also accessible with a precision superior to that of the LHC~\cite{Sun:2018gqo,Alvarez:2017ybk}. Competitive measurements of $V_{td}$ and $V_{ts}$ could be performed at the LHC with O(1)\,ab$^{-1}$ of integrated luminosity.

In addition, important measurements of other top-quark properties, such as of the top-quark spin and polarisation, can be performed with competitive precision compared to the LHC, but with the advantage of theoretical simplicity and experimental clearness~\cite{Atag:2006by}.

The photo-production of $t\overline{t}$ provides a window of opportunity to measure the $t\overline{t}\gamma$ magnetic and electric dipole moments~\cite{Bouzas:2013jha}. Here an energetic photon couples only with the top quark so the cross-section depends directly on the $t\overline{t}\gamma$ coupling. The sensitivity of the LHeC here is superior to that achieved with measurements of the $b\rightarrow s\gamma$ transition and with the production of $t\overline{t}\gamma$ at the LHC.

The LHeC also provides access to Flavor Changing Neutral Current (FCNC) processes driven by the $\gamma tq$ and $Ztq$ vertexes, where $q=u,c$~\cite{TurkCakir:2017rvu,Behera:2018ryv}. This is achieved by measuring the process $e^-p\rightarrow e^-W^{\pm}q +X$. The expected sensitivity improves on current limits from the LHC by up to one order of magnitude in case of the $\gamma tu$ coupling, and is competitive with the expected accuracy from the HL-LHC. 

\subsection{Precision Standard Model Physics}
Through its Lagrangian, the Standard Model of particle physics
defines all fundamental interactions  of the
elementary particles.
While the SM is a highly successful theory, a
main limitation of its exploration is that relevant free parameters,
especially the
coupling constant of the strong interaction $\alpha_s(M_Z)$ and the
weak mixing angle, $\sin^2\theta_W$, have only been
measured with moderate precision.
Any interference with new physics, from e.g. 
the dark sector, is expected to modify  loop
corrections of the SM, altering the SM predictions with
increasing scales.
Hence, the precise measurement of the fundamental parameters of the SM
including their scale dependence is of crucial importance.

\subsubsection*{Strong Coupling Constant}

The value of $\alpha_s(M_Z)$ is
determined at the LHeC through  precision measurements of scaling violations
and as well from jet measurements in the Breit frame. Highest
experimental precision is achieved since a color-neutral probe (the
lepton) scatters off a colored parton inside the proton, which provides
direct access to QCD phenomena and highest experimental precision
through the charged lepton in the final state.
The inclusive measurement over a large kinematic range in $Q^2$
and $x$
requires precise vertexing and the measurement of the
energy-momentum balance along the beam axis~\cite{Bassler:1994uq,Bassler:1997tv},
aligned with the requirements to measure PDFs as listed above.

These measurements are
improved with the challenging Silicon tracking detectors, here extended, and highly
granular calorimeters with complete azimuthal angular acceptance.
When determined together with the PDFs, the
value of $\alpha_s(M_Z)$ will be determined at the LHeC with an uncertainty of
$\pm0.00022$~\cite{Agostini:2020fmq}. This represents an impressive improvement
by a factor of five as compared to the present world average
value~\cite{PDG2021}. 

The measurement of inclusive jet cross sections is known to provide 
direct access to $\alpha_s(M_Z)$.
These jet  measurements require as well precise tracking,
vertexing and a high resolution calorimeter with efficient
discrimination between electromagnetic and hadronic clusters~\cite{Schwanenberger:2002pp,Kogler:2011zz}.
With its highly granular calorimeter, precise tracking, and the
in-situ calibration (see below), the jet-energy scale uncertainty at
the LHeC will be as small as about 0.5\,\%. 
This will result in a determination of $\alpha_s(M_Z)$ with an
experimental uncertainty of $\pm0.00016$~\cite{Agostini:2020fmq}, 
which is about a factor 10 more
precise than the best measurements from HERA jet data~\cite{Andreev:2017vxu,Britzger:2019kkb}.

A per mille level measurement of $\alpha_s$  eventually 
challenges the so far dominating lattice result. It
will pose new demands for theoreticians requiring to match that
small experimental uncertainty with improved higher-order predictions in QCD.
The equally precise determination in inclusive and jet production in DIS is 
a crucial cross check for the theory, and so will be a comparison to
future $e^+e^-$ precision measurements. It is remarkable that 
this result will be free of higher twist and nuclear corrections.
The LHC is unable to reach such precision
but needs it, as for example for the interpretation of the gluon-gluon fusion
production of the Higgs Boson being $\propto \alpha_s^2$.

At the LHeC the scale dependence of the strong coupling
can be determined in a large range in (renormalization) scale, from a
few GeV up to several hundreds of GeV, with small
uncertainties. Such a unique measurement provides a direct test of the
SU$_c$(3) gauge-structure of the strong sector of the SM.

\subsubsection*{Electroweak Measurements}

The space-like momentum transfer in deep-inelastic scattering is
mediated by electroweak interactions.
With increasing $Q^2$ the contributions to NC scattering from $\gamma Z$ interference 
and pure $Z$
exchange become more important, 
while CC DIS is mediated exclusively by a weak $W$ boson.
This provides the opportunity for direct measurements of fundamental
properties of the electroweak sector of the SM.
A highly granular electromagnetic calorimeter (EMC) improves the
measurement through identification of final-state QED photons. 
At the LHeC the weak mixing angle will be measured through polarisation  asymmetry and NC/CC ratio measurements in $e^-p$ scattering as has been
discussed in the initial CDR of the LHeC~\cite{AbelleiraFernandez:2012cc}.
With an uncertainty of $\Delta\sin^2\theta_W=\pm0.00022$~\cite{Britzger:2020kgg}
this is per se not an improvement over the precise $Z$-pole
measurements at LEP+SLC~\cite{ALEPH:2005ab}.
DIS, however, is mediated in the $t$-channel and thus the
scale dependence of $\sin^2\theta_W$ can be measured for the first
time with percent precision in the range of 
$\sqrt{Q^2}$ between $20$\,GeV up to nearly a TeV~\cite{Agostini:2020fmq}.
This
could
represent the first precision measurement of the weak mixing angle at
high scales, away from the $Z$-pole, an inevitable measurement for
particle physics, and a direct test of the gauge structure of the weak
sector of the SM.

The $Z$-exchange directly probes the weak NC couplings of the light
quarks inside the proton ($u$,$d$) to the $Z$-boson, and uncertainties
of the vector and axial-vector couplings of the light-quarks of about
1\,\% are achieved, which is an improvement by a factor of ten and more as
compared to present best measurements~\cite{PDG2021,Spiesberger:2018vki}.

The unique precision of the PDF and $\alpha_s$ measurements in $ep$
are a requirement to measure the mass of the $W$ boson to a few MeV 
precision, and similarly to extract a world leading measurement of 
the mixing angle from the LHC $pp$ measurements as is discussed 
in~\cite{Agostini:2020fmq}.

\subsection{Beyond the Standard Model Searches}
The clean environment of high-energy electron-hadron collisions provides an excellent framework for studying many extensions of the Standard Model.
The excellent detector performance, the absence of pileup, and the large luminosity allows testing of entire classes of models that are difficult to study at the LHC.
Many studies from recent years have been summarised succinctly in chapter 8 of ref.~\cite{Agostini:2020fmq}.

Prominent examples among these studies are searches for sterile neutrinos, for instance via lepton-trijets and displaced vertex signatures \cite{Antusch:2019eiz}, heavy scalar particles with masses around the electroweak scale \cite{DelleRose:2018ndz}, and in general models with final states that look like `hadronic noise' in proton-proton collisions \cite{Curtin:2017bxr,Curtin:2018xsc}.

Recent studies demonstrate that the LHeC could be a world-leading laboratory to study flavor-changing neutral currents in the charged lepton sector, in particular for processes that lead to electron-to-tau transitions, where the projected sensitivity could be an order of magnitude better than current and planned experiments in tau factories \cite{Antusch:2020vul}.

Scalar and fermion $SU(2)_L$ triplets can explain the observation of neutrino masses via the so-called type-II and type-III mechanisms, respectively. Both types of particles can be produced via their gauge interactions in vector boson fusion, but studying them at the LHC is very challenging due to the towering backgrounds.
The prospects of finding triplet fermions via fat jet final states were shown to be feasible at the LHeC \cite{Das:2020gnt}.
Triplet scalar searches at the LHeC were discussed in ref.~\cite{Yang:2021skb}.

Certain classes of leptoquarks can be studied at the LHeC if they interact with first generation fermions and have decay channels that are difficult to reconstruct at the LHC. It is possible to test  explanations of the flavor anomaly $R_{D^{(*)}}$ via the $R_2$ leptoquark at the LHeC via its decays into $\tau b$ final states \cite{Azuelos:2020gvi}.

Less minimal models with a $\tilde R_2$ leptoquark that has a dominant branching ratio into right-handed neutrinos may escape the LHC searches, but can be studied at the  LHeC~\cite{Padhan:2019dcp}. The specific signature of a displaced fat jet, stemming from the decay of a long lived heavy neutrino, would be a very promising sign of this model at the LHeC \cite{Cottin:2021tfo} and could already be observable within the first few months of operation.

Dark photons with masses below 10 GeV can be tested in a decay-agnostic approach via distinct non-dglap scaling violations, which may be the smoking gun for LHeC searches \cite{Kribs:2020vyk}. In the event that the dark photon in this mass range is long lived and decays dominantly to lepton pairs, LHeC searches for displaced dark photon decays would be sensitive to an otherwise challenging region of the parameter space \cite{DOnofrio:2019dcp}.

BSM physics poses requirements on the detector design which
are important to consider for the tracker, calorimeter and muon detectors, 
such as an impact parameter resolution of $10$\,$\mu$m,
very good momentum resolution, low mass photon
identification, high resolution 
missing energy, detection of long lived particles outside the inner tracker, full
hermiticity and maximum pseudo-rapidity coverage to $\eta \sim 5$.

\section{Detector Design}
\label{detector}

The LHeC detector, initially designed in~\cite{AbelleiraFernandez:2012cc} and recently updated~\cite{Agostini:2020fmq}
as a modern general purpose $ep$ detector, is a composite system made of several subcomponents: beampipe, tracking, calorimetry, magnets and a muon system, each optimised for its purpose and adapted to the interaction region, which has the peculiarity of hosting 3 beams, the 2 proton or ion beams, of which one is a spectator while the other one is interacting with the counter-rotating electron beam, or, in the joint $eh/hh$ configuration possibly the hadron beam. 

In the following section the LHeC detector baseline design and some of its subcomponents are discussed illustrating few aspects and recent developments. Some consideration to adapt the detector for higher energy running (HE-LHC or FCC)  are briefly addressed.  
A separate Section~\ref{subsec:ehandhh} below is devoted to modifications that would be required should the LHeC detector be used to also register hadron-hadron collisions.

\subsection{Requirements}
\label{section-det-requirements}

Requirements on the LHeC detector have been presented above concluding the
discussions of the key physics topics.
The detector should be highly hermetic in order to maximise coverage, in both the forward and backward directions\footnote{Positive $z$ is defined as the incoming hadron beam direction for $ep/eA$ collisions.}, to provide a precise measurement of the hadronic final states and of scattered electrons towards very low-$Q^2$. For charged current processes, the reconstruction of kinematic variables is only possible through the hadronic final state measurement where an excellent performance on calorimetry for hadrons is required to reconstruct the missing energies and distinguish CC from NC events
at low $Q^2$.
The good hermeticity is also important for calibration of the detector through transverse momentum balance using NC DIS and photo-produced di-jet events.

Fine segmentation and good resolution for the electromagnetic calorimeter is required all over the angular coverage to tag both low-$Q^2$ and high-$Q^2$ neutral current events. Good resolution in the hadronic section is also important to measure the missing energies for CC DIS as well as for QCD studies using jets.

Excellent flavour tagging performance is desirable, especially in the forward direction, for flavour decomposition of jets and for tagging the SM Higgs decaying to $b\bar{b}$ and $c\bar{c}$, which are predominantly produced \textcolor{red}{at} large positive $\eta$.

There are also various constraints and considerations to take into account from the accelerator and technical aspect of detectors:
\begin{itemize}

\item The detector is required to have a magnet system consisting of a central solenoid 
along with a dipole system to steer the electron beam allowing for head-on $eh$ 
collisions at the interaction point;

\item 
The non-interacting proton/ion beam has to bypass the $ep$ interaction yet to be guided through the same beam pipe housing the interacting electron and proton/ion beams; 

\item The shape of the beam pipe has to allow for the electron beam generated synchrotron fan
to leave the interaction region unaffected and with minimal back-scattering; 

\item Good vertex resolution 
implies a small radius and thin beam pipe optimised in view of synchrotron radiation and background effects;

\item The tracking and calorimetry in the forward and backward directions are set up to take into account the extreme asymmetry of the DIS production kinematics, see ~\cite{AbelleiraFernandez:2012cc}, with multi-TeV energies
emitted in the forward, proton beam direction while the
electromagnetic and hadron energies emitted backwards are
limited by the electron beam energy.

\item Very forward and backward detectors have to be set up to access diffractive produced events and to tag  photo-production processes besides measuring the luminosity with high precision in Bether-Heitler scattering, respectively.

\end{itemize}
These and further specific requirements
from inclusive DIS, see Sect.\,\ref{sec:partons}, are basically known from the H1 and ZEUS experiments at HERA. However, at the LHeC they are posed with
extra severity because of the much enlarged beam energies, broad physics programme and more ambitious precision demands driven by new physics opportunities and owing to the huge increase in luminosity - as a few days of LHeC operation correspond to 15 years of HERA operation. New requirements are worth emphasising that
arise from a much extended physics programme for which Higgs measurements are a prime example.

\subsection{A Detector for DIS at the LHC}

\begin{figure}[htb]
\includegraphics[width=0.49\textwidth]{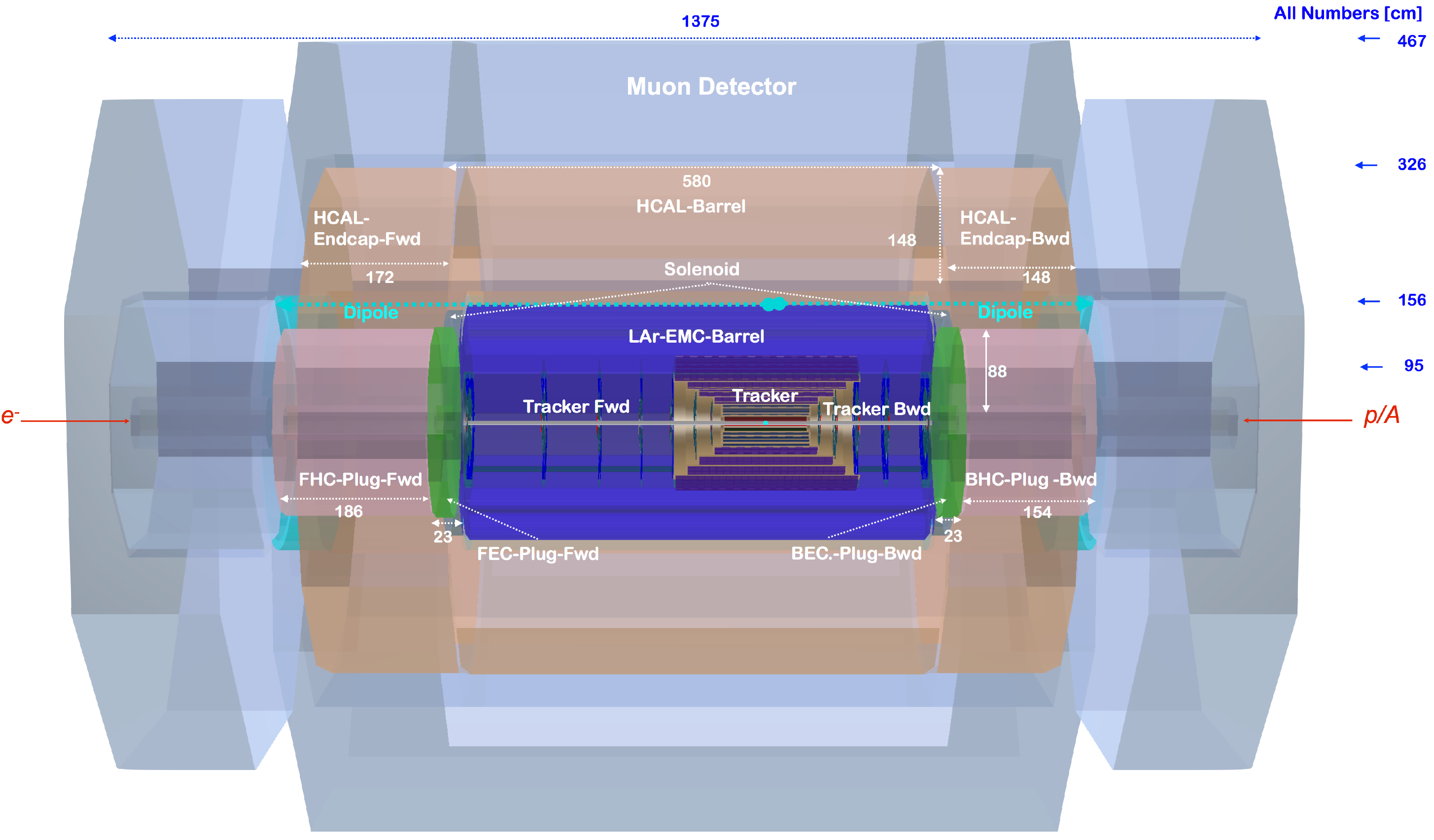}
\caption{Side view of the updated baseline LHeC detector concept, providing an overview of the main detector components and their locations. The detector dimensions are about 13m length and 9m diameter. The tracker is setup using pixel, macropixel and strip detectors. The barrel elctromagnetic LAr-calorimeter EMC (in blue) surrounding the tracking region. The solenoid magnet is placed at radii immediately outside the EMC-Barrel, and is housed in a cryostat, which it shares with the weak dipole magnet that ensures head-on collisions. The hadronic calorimeter HCAL in the barrel part (colored orange; it uses steel\,$\&$\, scintillating tiles) is located outside of the solenoid. The forward/backward electromagnetic calorimeters FEC/BEC (in green) and hadronic calorimeters FHC/BHC (in bright orange) are using Si-based sensitive\,$\&$\,readout technology and as absorbers W/Pb and W/Cu, respectively~\cite{AbelleiraFernandez:2012cc,Agostini:2020fmq}. The muon detector (in grey) forms the outer shell of the detector. 
The detector description has been setup using {\small\bf{DD4hep}}~\cite{frank_markus_2018_1464634}.}
\label{fig:det:sideview}
\end{figure}

The present LHeC detector is illustrated in  Fig.~\ref{fig:det:sideview}. The detector is asymmetric in design, reflecting the beam energy asymmetry.
The design is largely based on established technologies from the LHC general purpose detectors, ATLAS and CMS. More advanced technologies are introduced, and will evolve with time, to fulfill the above described requirements and to adapt to different running conditions. The detector covers the angular range from $1^\circ$ to $179^\circ$ by the calorimeters to achieve the required hermeticity. Compared to $pp$ running, the expected $ep$ collision and background rates are about three orders of magnitude smaller relaxing the requirements on radiation hardness and also data acquisition. The pile-up rate is less than 0.1 per crossing at the LHeC for $10^{34}$\,cm$^{-2}$s$^{-1}$. 
The neutron field is expected to be a few orders of magnitude smaller than the LHC environment, estimated to be about $10^{13}$neq/cm$^2$. 

As illustrated below in Section~\ref{ElectronBeamOptics} (cf. Fig.~\ref{fig:sep_scheme}) a dipole field is needed to steer the electron beam in the interaction region and allow for head-on collisions with the proton beam. The required dipole field (0.17\,T over the range $z=[-8$m, $+8$m$]$) is combined in the central region with the central solenoid providing a field of 3\,T, for a technical design see~\cite{AbelleiraFernandez:2012cc}. The synchrotron radiation generated by the electron beam in the dipole field is leaving the interaction region not affecting the detector performance thanks to the asymmetric design of the lightweight Beryllium beampipe. 

The main detector, from the interaction point to the outer radius, consists of: the silicon tracker with a central barrel part, forward and backward wheels; the electromagnetic calorimeter housed inside solenoid and dipole magnet; the hadronic calorimeter and the muon system. Not shown in the figure are backward (electron-side) detectors for low-angle scattered electron to tag $\gamma p$ and $\gamma A$ collisions and forward detectors for neutrals ($n$, $\pi^0$ ...) from the $p/A$ remnant and protons spectrometer to measure proton momentum from elastic and quasi-elastic scattering.

This baseline design serves also a scalable configuration for HE-LHC and FCC-he where the main changes to be made for higher collision energy are the extensions for rapidity coverage in the tracking system and the depth ($X_0, \lambda_I$) in the calorimetry, both affecting mainly the size of the detector in the beam direction but only logarithmically.
With respect to the earlier versions presented in the CDR and in the update, some optimisation has been done in particular to the silicon tracking, extending 
the radius from 60 to 80\,cm, and to the calorimetry, choosing LAr fr the EMC, which are described below in more detail. The larger tracking volume, with a longer lever arm measurement using more track points, 
allows for better resolution even at a slightly reduced B field. 
We expect that this configuration will deliver good and stable performance also in different experiment and accelerator configurations ($eh$ and $hh$ running) with certain modifications for $hh$ as  are described in Section~\ref{subsec:ehandhh}.

\subsubsection{Silicon Tracking System}
\label{LHeC_tracker}
\begin{figure}[htb]
\includegraphics[width=0.49\textwidth]{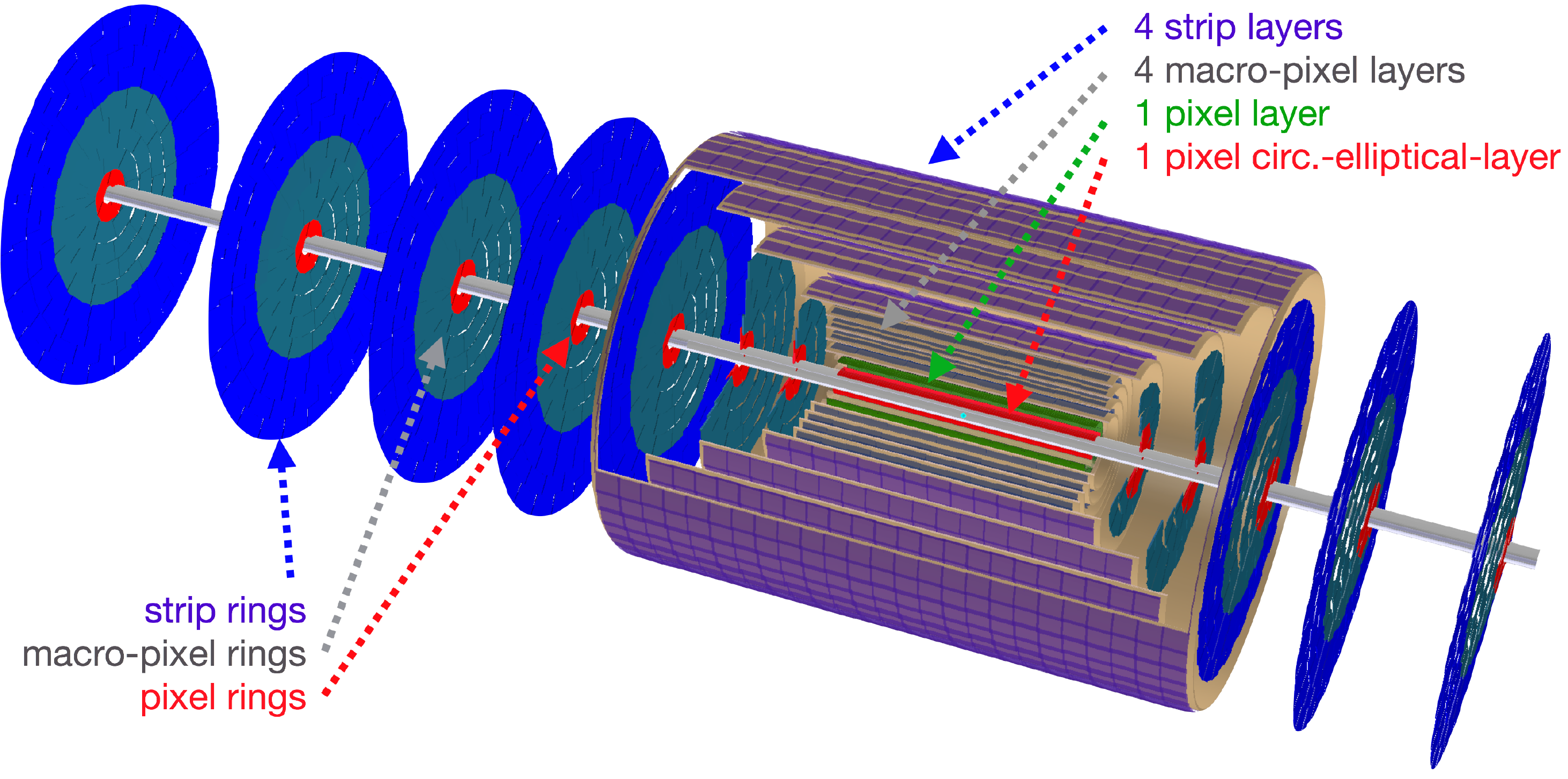}
\caption{Structure of the  Silicon  tracker for the LHeC detector.}
\label{fig:det:SiTracker_fwd_bwd}
\end{figure}

\begin{figure}[htb]
\begin{center}
\includegraphics[width=0.32\textwidth]{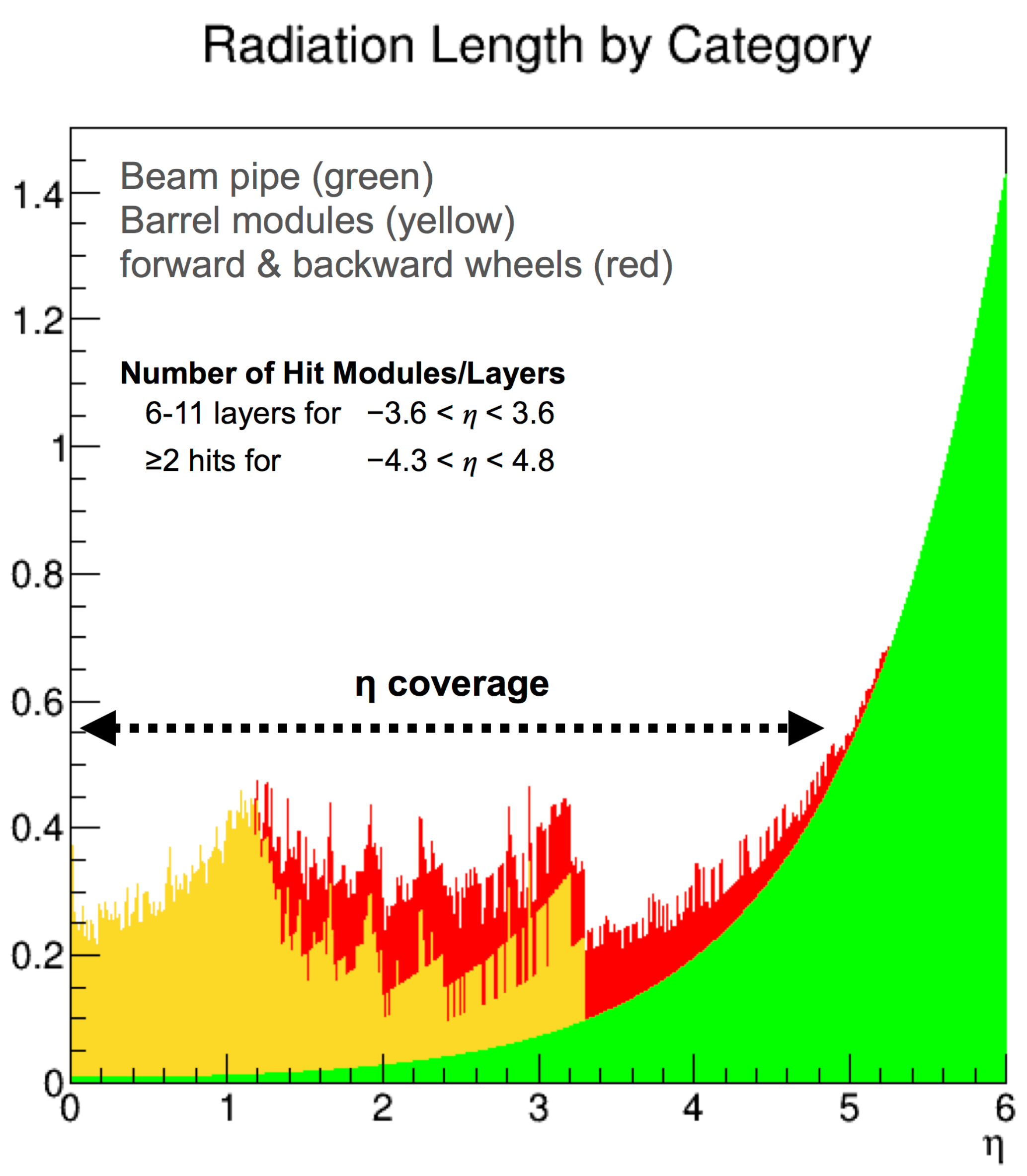}
\end{center}
\caption{Calculation of the radiation length budget of the LHeC tracker using {\scriptsize\bf{tkLayout}}~\cite{Bianchi:2014mpa}. Support structures and services are not included.}
\label{fig:det:SiTracker_radiation-length_hit-coverage.pdf}
\end{figure}

As described previously, excellent flavour tagging ability, including charm decays, is required across wide angular range, in particular towards forward rapidities. The decay particles from the SM Higgs may go beyond $|\eta| > 2.5$, the usual tracking coverage for the LHC $pp$ detectors. The silicon tracker is shown in Fig.\,\ref{fig:det:SiTracker_fwd_bwd}. It covers up to $|\eta| < 3.6$ with at least six hits and two hits for $-4.3 < \eta < 4.8$ using seven forward and five backward disks. In comparison to the previous LHeC tracker, the outer radius was extended from 60\,cm to 80\,cm and the number of layers in the barrel region from 7 to 10 layers while the magnetic field of the solenoid was reduced from 3.5\,T  to 3\,T. 
Using the {\small\bf{tkLayout}} tool\,\cite{Bianchi:2014mpa} for optimising the tracker arrangement and minimising the material impact over a large region of $\eta$, the calculated radiation length figure shows tolerable levels as shown in Fig.~\ref{fig:det:SiTracker_radiation-length_hit-coverage.pdf}. 
Some properties of the tracker setup are summarised in Tab.\,\ref{tab:det:LHeC_Tracker_main-properties}.

\begin{table}[ht]
\renewcommand{\arraystretch}{0.8}
\label{tab:det:LHeC_Tracker_main-properties} 
\begin{tabular}{|l||c|c|c|}
\hline     
\scriptsize{\textbf{LHeC Tracker Part}} \hfill &
\scriptsize{$\eta{_\text{max}}$ }    & 
\scriptsize{$\eta{_\text{min}}$ }    & 
\scriptsize{{\#}Layers${_{\tiny{\textrm{Barrel}}}}$} \\
\hline
\hfill \scriptsize{pix}           & \scriptsize{3.3} & \scriptsize{-3.3} & \scriptsize{2} \\
\scriptsize{\textbf{Inner\,Barrel}} \hfill  \scriptsize{pix$_{\textrm{macro}}$}  & \scriptsize{2.} & \scriptsize{-2.} & \scriptsize{4} \\
\hfill \scriptsize{strip}         & \scriptsize{1.3} & \scriptsize{-1.3} & \scriptsize{4} \\
\hline
   & & & {\scriptsize{{\#}Rings${_{\tiny{\textrm{Wheels}}}}$} } \\
\hline
\hfill \scriptsize{pix}           & \scriptsize{4.1/-1.1} & \scriptsize{1.1/-4.1} & \scriptsize{2} \\
\scriptsize{\textbf{End Caps}} \hfill  \scriptsize{pix$_{macro}$}  & \scriptsize{2.3/-1.4}   & \scriptsize{1.4/-2.3} & \scriptsize{1}  \\
\hfill \scriptsize{strip}         & \scriptsize{2./-0.7} & \scriptsize{0.7/-2.} & \scriptsize{1-4} \\
\hline
\hfill \scriptsize{pix}           & \scriptsize{5.2} & \scriptsize{2.6} &  \scriptsize{2} \\
\scriptsize{\textbf{Fwd\,Tracker}}  \hfill  \scriptsize{pix$_{\textrm{macro}}$}  & \scriptsize{3.4}   &  \scriptsize{2.2} &  \scriptsize{1} \\
\hfill \scriptsize{strip}         &  \scriptsize{3.1}   & \scriptsize{1.4} &  \scriptsize{4}  \\
\hline
\hfill \scriptsize{pix}           & \scriptsize{-2.6} & \scriptsize{-4.6} & \scriptsize{2} \\
\scriptsize{\textbf{Bwd\,Tracker}}  \hfill  \scriptsize{pix$_{\textrm{macro}}$}  & \scriptsize{-2.2} & \scriptsize{-2.9} & \scriptsize{1} \\
\hfill \scriptsize{strip}         & \scriptsize{-1.4} & \scriptsize{-2.5} & \scriptsize{4}  \\

\hline

\scriptsize{Total~$\eta{_\text{max/min}}$ } \hfill    & \scriptsize{5.2}  & \scriptsize{-4.6} & \\
\hline
\end{tabular}
\caption{Summary of the main properties of the tracker modules in the revised LHeC detector configuration based on calculations performed using tkLayout~\cite{Bianchi:2014mpa}.
$\eta{_\text{max/min}}$ denotes the pseudo-rapidity range. {\#}Layers${_{\tiny{\textrm{Barrel}}}}$  are the number of layers in the barrel and {\#}Rings${_{\tiny{\textrm{Wheels}}}}$ the number of wheels in the End Caps, Fwd and Bwd tracker parts, respectively.}
\end{table}

The relatively small radiation level allows to employ CMOS-based technology for the inner silicon tracker. Depleted CMOS sensors, also known as Depleted Monolithic Active Pixel Sensors (DMAPS), are position sensitive detectors in industry standard CMOS or High Voltage-CMOS (HV-CMOS) processes~\cite{Peric:2007qtk}. These sensors are extremely attractive for experiments in particle physics as they integrate the sensing element and the readout electronics in a single layer of silicon, which removes the need for interconnection with complex and expensive solder bump technology. Depleted CMOS sensors also benefit from faster turnaround times and lower production costs when compared to hybrid silicon sensors. The final choice will depend on the region of exploration. Low-fill factor DMAPS have been or are being prototyped and produced for several experiments in particle physics, such as Mu3e~\cite{Blondel:2013ia}, ATLAS~\cite{Gabrielli:2018qdl}, LHCb~\cite{Ackernley}, CLIC~\cite{Linssen} and ALICE~\cite{Adamova:2019vkf,Preghenella:2020mxn} in a few different processes. Today’s most performant DMAPS detectors are 50$\mu$m thin and have {50$\mu$m x 50$\mu$m} cell size with integrated mixed analogue and digital readout electronics, 6ns time resolution and {$2 \times 10^{15}$ 1MeV neq/${\rm cm^{2}}$} radiation tolerance. The development is ongoing and extends towards radiation hard technologies. Interesting for our purpose are the possibilities of features offered by CMOS imaging sensor technologies, called stitching, which allows developing a new generation of larger size MAPS  using wafers that are 300mm in diameter. 
Moreover, the reduction of the sensor thickness to values as small as about 20-40$\mu$m shall allow for exploiting the flexible nature of silicon to possibly implement large-area curved sensors. In this way, it looks feasible to build cylindrical or in general curved layers of silicon-only sensors, with a significant reduction of the material thickness by avoiding overlap between sensors~\cite{ALICE_LS3,Mager,Rinella:2021iyw}.

\begin{figure}[htb]
\begin{center}
\includegraphics[width=0.45\textwidth]{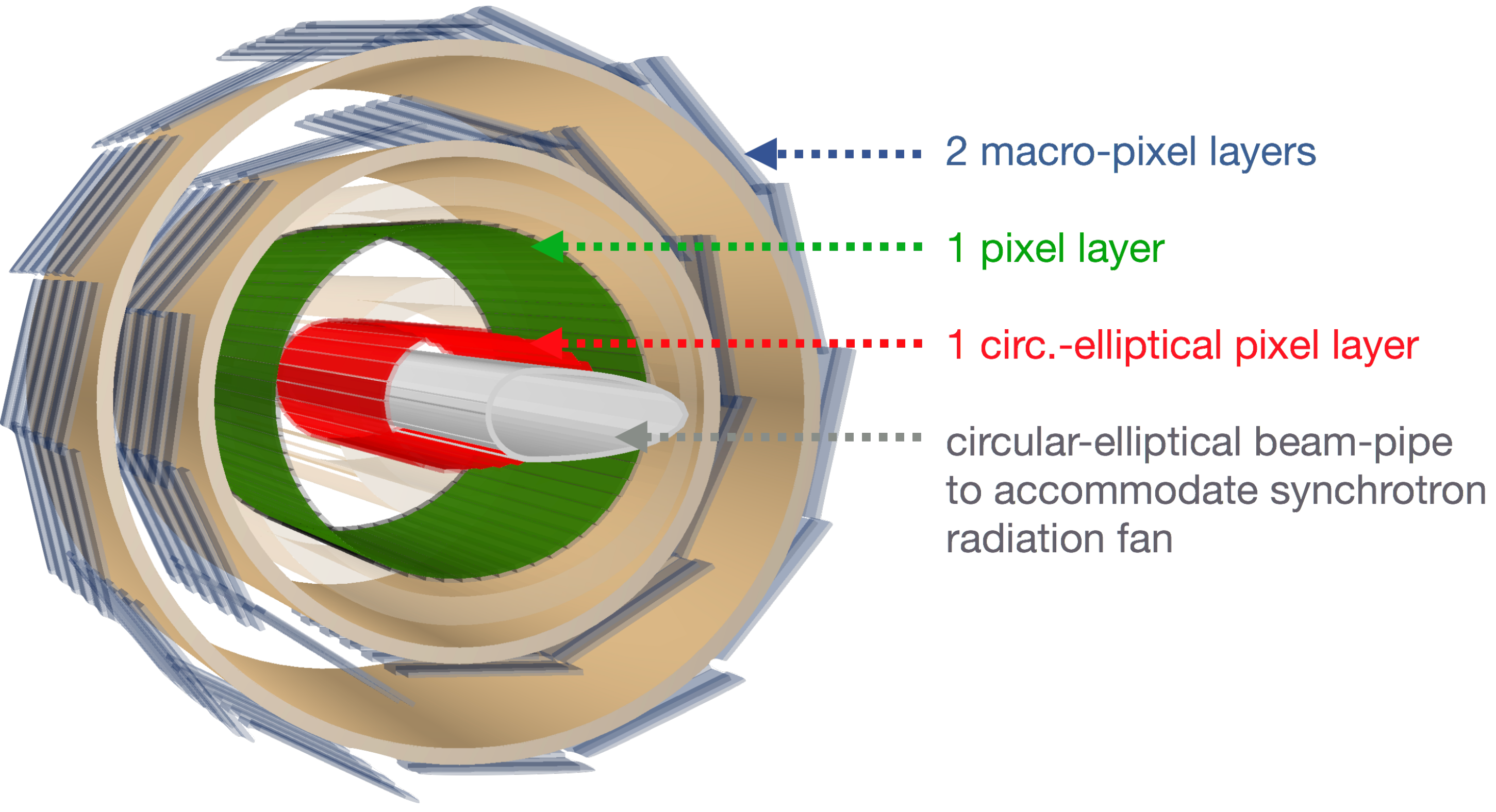}
\end{center}
\caption{A view of inner 4 layers of central barrel tracker with innermost circular-elliptical silicon pixel layer following the shape of beampipe.}
\label{fig:det:innermostLayers}
\end{figure}

The challenge in vertexing at the LHeC is that the beampipe has  to accommodate the synchrotron radiation fan from the electron beam. To minimise the impact, the innermost barrel pixel layer is designed to follow the \textcolor{red}{\sout{an}} optimized circular-elliptic shape of the beampipe as shown in Fig.\,\ref{fig:det:innermostLayers}. Thanks to the integrated read-out electronics of the DMAPS sensors, the layout of the innermost layer can be flexible. The current design uses a scheme with many narrow sensors in the x\,-\,y coordinate plane, see Fig.\,\ref{fig:det:innermostLayers}, following the shape of beampipe as closely as possible.
A possibility to use the bent sensors as described above is being studied.

\subsubsection{Calorimetry}
\label{sec:calo}
As described in Section \ref{section-det-requirements}, the LHeC requires well developed electromagnetic and hadronic calorimeter sections. The electromagnetic calorimeter (EMC) surrounds comple\-tely the silicon tracker and can be subdivided into a barrel, a forward and a rear system. 

For the barrel region two options had been 
considered: a cold option using Liquid Argon, copper electrodes and lead absorbers, and a warm one based on lead absorbers and scintillator tiles. Liquid Argon is known for its high resolution, linearity, long term stability and radiation tolerance confirmed over many years in ATLAS and H1 experiments~\cite{Babaev:1994pd,Abt:1996hi,Abt:1996xv,Fleischer:1997aq,Issever:2000yh,Schwanenberger:2002pp,Seehafer:2005ma,Kiesling:2010zz,Aubert:2005dh,ATLAS:1996ab,Morgenstern:2019xto}. The cryogenic system required for the LAr option can be combined in the LHeC detector with the one from the magnet system, which is directly surrounding the calorimeter. The flexibility in the longitudinal and transverse
segmentation, and the possibility of implementing a section with narrow strips to measure the shower shape in its initial development, represent additional advantages~\cite{ATLAS:1996ab}.

\begin{figure}[htb]
\includegraphics[width=0.49\textwidth]{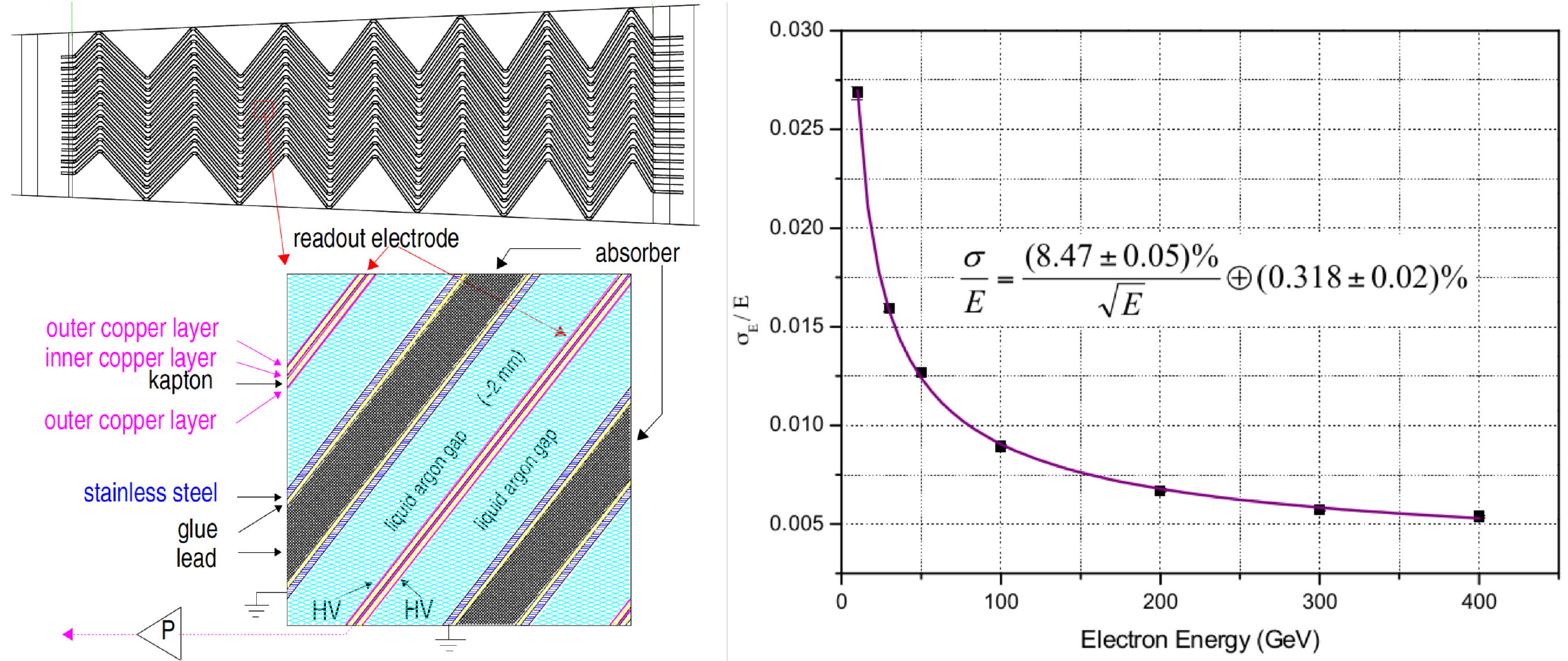}
\caption{Longitudinal view of one cell of the ATLAS LAr Calorimeter,
  showing the accordion structure (left). The LHeC LAr energy resolution for electrons between 10 and 400 GeV as simulated using GEANT4~\cite{Agostinelli:2002hh} (right)~\cite{AbelleiraFernandez:2012cc}. }
\label{fig:det:LAr_cell_resolution}
\end{figure}

Fig.~\ref{fig:det:LAr_cell_resolution}\,(left) shows a detail of the accordion-electrode structure.  A basic cell consists of an absorber plate, a liquid argon gap, a readout electrode and a second liquid
argon gap. The mean thickness of the liquid argon gap is constant along the whole barrel and along the calorimeter depth. The LHeC LAr electromagnetic calorimeter would also provide the required energy resolution and detector granularity (Fig.\,\ref{fig:det:LAr_cell_resolution}\,(right)).
More details on the LAr calorimeters proposed can be found in the CDR~\cite{AbelleiraFernandez:2012cc}. As an option, in~\cite{Agostini:2020fmq},
alternatively a warm
lead-scintillator electromagnetic calorimeter has been simulated for comparison. Its advantage compared to the LAr calorimeter are no cryostat walls in front of the barrel EMC and a convenient modularity. Given, however,
its significantly better resolution performance, preference was decided to give 
to a barrel LAr calorimeter.

The hadronic calorimeter in the barrel part is a sampling calorimeter using steel and scintillating tiles as absorber and active material, respectively, for good resolution. This also provides mechanical stability for the magnet/dipole cryostat and the tracking system.

Calorimetry in the forward and backward direction at the LHeC needs very fine granularity for position resolution, good $e/\pi$ separation through shower shape and also good resolution, especially for the scattered electron.
The very forward and to a much lesser extent -- in $eh$ -- the backward parts of the calorimeter are exposed to high levels of particle radiation and must therefore be radiation hard by design. Tungsten (W) is considered as the absorber material, in particular for the forward inserts (electromagnetic and hadronic inserts), because of its very short radiation length. Since the backward inserts have looser requirements, the materials for the absorbers are chosen as lead (Pb) for the electromagnetic part and copper (Cu) for the hadronic one. 

The active signal sensors have been chosen to be silicon-strip for the electromagnetic forward/backward calorimeters and silicon-pad for the hadronic fwd/bwd calo\-ri\-me\-ters. The demanding requirements of very for\-ward /backward angle resolution favors fine segmentations of calorimeter cells interconnecting the tracking and calorimeter information for best particle-tracking and -identification. Si based tracking/imaging calorimeters   
appear to be appropriate to withstand the higher radiation load near the beam-pipe. They also open the opportunity to measure the neutral component of particle flow as already demonstrated by developments of the  CALICE collaboration for the Linear Collider\,\cite{Boudry:2020dov,Cabrera:2021vih,Kawagoe:2019tno}. 

The hadronic calorimeter compensation algorithm would profit as well from knowing the neutral part of shower development best.
The steel structures are in the central and plug calorimetry close the outer field of the central solenoid. 
The total depth of the electromagnetic section is about 30\,radiation lengths on average in the barrel and backward regions. In the forward direction, where particle and energy densities are highest, the segmentation/granularity has to be finer, and it varies with radius and depth. The hadronic calorimeter has a depth of between 7.1 and 9.6\,interaction lengths,
with the largest values in the forward plug region. For each of the calorimeter modules, the pseudorapidity coverage, the types of the absorber and sensitive materials used, the number of layers, radiation or interaction lengths, and the energy resolutions obtained from GEANT4 simulations can be found in Ref.~\cite{Agostini:2020fmq}.

\subsubsection{Muon System}

Muon identification is an important aspect for any general purpose HEP experiment. 
In the baseline LHeC detector design, the muon system provides a reliable muon tag signature which is used in conjunction with the central detector for muon identification, triggering and precision measurements. Detection of long lived particle decays suggests to
foresee not too short  track segment lengths in the muon detector.
The detector elements are organised in a nearly hermetic envelope surrounding the hadronic calorimetry. In terms of technology choices, the options in use in the LHC general purpose experiments~\cite{atlas-phase2-upgrade-tdr,cms-phase2-upgrade-tdr}
and their planned upgrades are adequate for LHeC  since muon background rates are lower. A solution composed of layers of last generation Resistive Plate Chambers (RPC), providing the Level 1 trigger and a two coordinate ($\eta$,$\phi$) measurement and possibly aided Monitored Drift Tubes for additional precision measurements appears as appropriate~\cite{Agostini:2020fmq}.
In the baseline design, the muon chambers have a compact multi-layer structure, providing a pointing trigger and a precise timing measurement which is used to separate muons coming from the interaction point from cosmics, beam halo and non prompt particles. This tagging feature does not include the muon momentum measurement, but is performed only in conjunction with the central detector. The muon detector will cover the region $|\eta|<4.1$ and the transverse momentum coverage can be roughly estimated to be $p_T> 2- 3$ GeV (at central pseudorapidities).


\section{Accelerator Considerations}
\label{machine}

The design of the machine is described in detail in the updated version of the LHeC design report~\cite{Agostini:2020fmq}.
It combines an optimised proton optics of the  LHC interaction region (IR 2)  with a dedicated new lattice for strong focusing of a \SI{50}{GeV} electron beam that will collide with one of the counter rotating proton beams. 

The electrons are accelerated in a new Energy recovery Linac (ERL)  structure that allows high beam currents and at the same time a moderate energy consumption.
The lattice of the combined interaction region and the beam optics of the three beams has to be flexible enough  to provide in alternative manner standard LHC beam operation of  $pp$ (or $AA$) collisions for heavy ion physics and electron-proton (or electron-ion) collisions at the same LHC interaction point ``IP2''.

The main challenges of this new layout are:
\begin{itemize}
    \item the design of a high intensity high energy ERL that will provide a \SI{50}{GeV} electron beam;
    \item an Interaction Region that is flexible enough for the alternative operation of $ep$ and $pp$ (or $AA$) collisions in IP2;
    \item and a beam combination and separation scheme to bring the electrons into head-on collisions with the \SI{7}{TeV}  proton beam (or 2.76 TeV/nucleon ion beam) and separate them after the IP.
\end{itemize}

Beyond these beam dynamics aspects,  the aperture need of the mini-beta concept of electron as well as proton beam are defined by the fact that the  $ep$ mode in IR2 will 
have to be combined with standard $pp$ collisions in the other LHC interaction points, IP1, IP5 and IP8.
As a consequence, the new interaction region is optimised to house three beams: the colliding electron and proton beams as well as the second proton beam that will pass  unaffected  the same vacuum system. A special relaxed beam optics is foreseen for this non-colliding protons to limit the additional aperture need.   
For the $pp$ operation mode - the standard scenario in present LHC operation - the so-called ATS optics will be applicable, that has been established in the context of the HL-LHC luminosity upgrade project, for both heavy ion and proton operation. 
The final focus system of the electron beam is embedded in this existing LHC proton structure.
As a consequence its mini beta focusing system has to be extremely compact to be nested in the proton mini beta lattice. It will provide transverse beam dimensions of the electrons to match in both planes the dimensions of the LHC hadron beam.  
At the same time the electron lattice has to provide a sufficient beam separation and guide the electrons after the collision point back to the ERL return arc for the deceleration part  of the energy recovery process.

\begin{figure}[!htb]
   \centering
   \includegraphics*[width=.9\columnwidth]{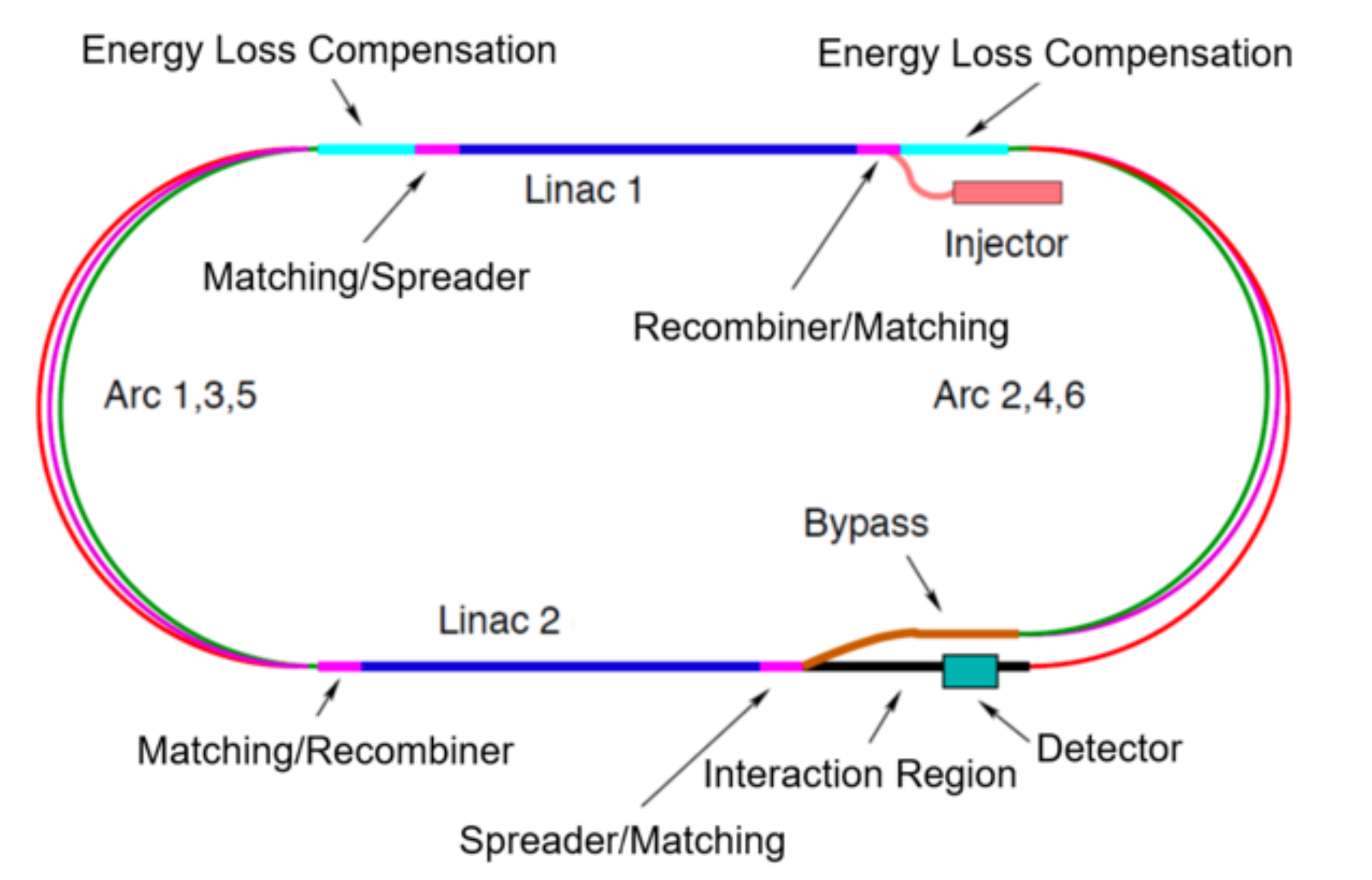}
   \caption{ERL geometry, using two superconducting linear accelerators, connected by three-pass return arcs.}
   \label{fig:ERL_sketch}
\end{figure}

 The ERL is based on two super-conducting linacs of about \SI{900}{m} length, which are placed opposite to each other and connected by three return arcs on both sides (Fig.~\ref{fig:ERL_sketch}). A final electron beam energy of  \SI{50}{GeV} is reached in this 3-turn racetrack ERL design. The concept allows to keep the overall energy consumption on a modest level for up to \SI{20}{mA} electron current.
 The main parameter list is shown in Tab.~\ref{tab:main_parameters}.
\begin{table}[htb]
\centering
      \begin{tabular}{l c c}
       \hline
       \textbf{Parameter} & \textbf{Unit}                       & \textbf{Value} \\
       \hline
         Beam energy & GeV &  50 \\
         Bunch charge & pC & 499 \\
         Bunch spacing & ns  & 24.95 \\
         Electron current & mA & 20 \\
         trans.~norm.~emittance &   $\mu$m & 30 \\
         RF frequency & MHz & 801.58 \\
         Acceleration gradient & MV/m & 20.06 \\
         Total length & m & 6665 \\
       \hline
      \end{tabular}
\caption{ERL main parameters}
    \label{tab:main_parameters}
\end{table}

\subsection{Linac and RF system}
\begin{figure}[!htb]
   \centering
   \includegraphics*[width=.8\columnwidth]{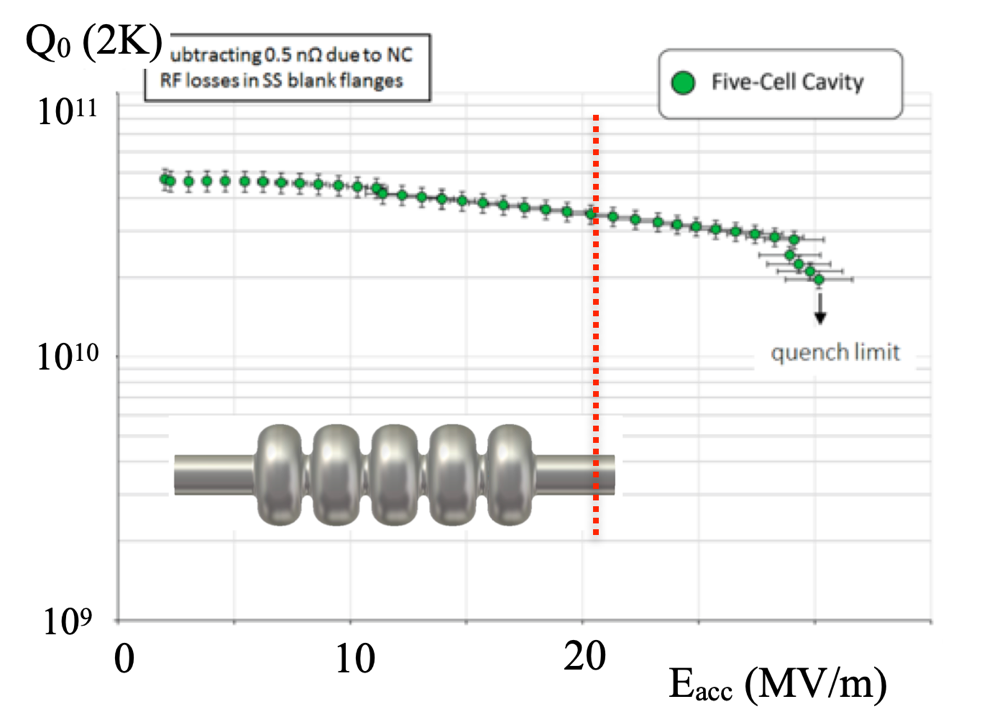}
   \caption{Q-parameter of the 5 cell cavity prototype}
   \label{fig:sc_cavity}
\end{figure}

The option to design a particle collider as Energy Recovery Linac, provides the  opportunity to overcome or avoid a number of  limitations of circular machines.  In order to reach the luminosity of \SI{e{34}}{\per \cm \squared \per \second} with an electron energy  of \SI{50}{GeV}, the concept of an ERL offers the advantage of a high brightness beam, high beam currents with limited synchrotron radiation losses and it avoids limitations due to the beam-beam effect - a major performance limitation in many circular lepton colliders (e.g. LEP). 
On the other side, the current of the ERL as well as the emittance are determined by its source. An operational goal of $I_e = $\SI{20} {mA} for the LHeC  has been set, corresponding to a bunch charge of 500\,pC at a bunch frequency
of 40 MHz. Given three turns for the acceleration and deceleration, an overall current of \SI{120}{mA}  will be circulating in the ERL with impacts on the RF design, facing a virtual beam power  of \SI{1}{GW}. In order to limit RF losses, a super conducting (s.c.) RF system is foreseen with a required quality factor above $Q=10^{10}$. In collaboration with JLab~\cite{J-labsccav} prototypes have been developed: Figure \ref{fig:sc_cavity} shows the Q-value of a five cell s.c. resonator which lies comfortably above this value up to the required acceleration gradient, which is indicated by the red cursor line in the plot, and 
much beyond.
The further SRF technology development, the validation of these design concepts and the investigation of the ERL  performance in terms of source brightness and  stable and efficient operation are important tasks of the PERLE facility~\cite{Angal-Kalinin:2017iup,PERLE}.
\subsection{Return Arcs and Spreaders}
\begin{figure}[!htb]
   \centering
   \includegraphics*[width=\columnwidth]{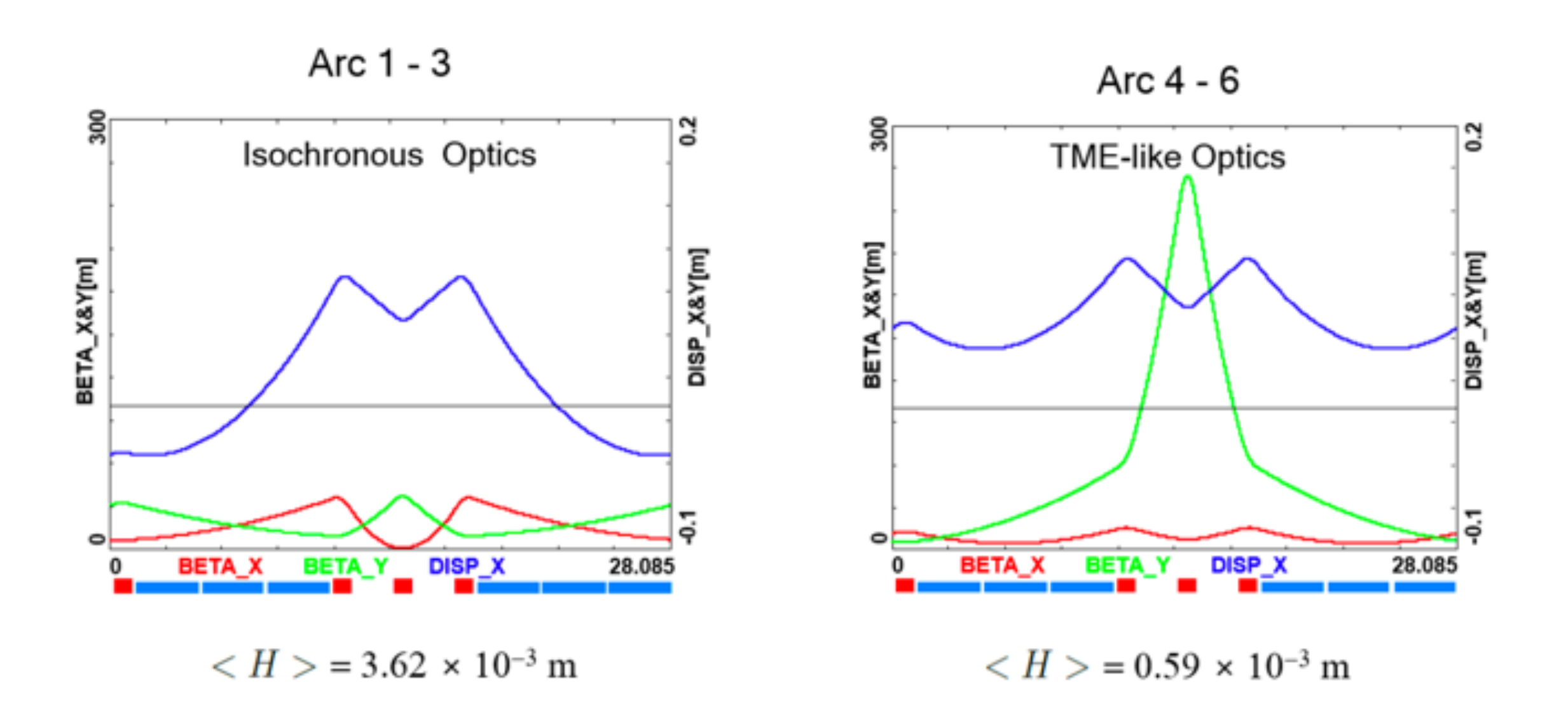}
   \caption{Basic FMC cells of the ERL arcs: Isochronous (left) for arc 1,2,3 and TME lattice (right) for arc 4,5,6.}
   \label{fig:arc_lattice_tme}
\end{figure}

Special care has to be taken in the design of the ERL lattice: The optics of the three return arcs  has to be optimised for the different challenges, that come along with the increasing beam energy~\cite{Alex}. At low energy, a flexible momentum compaction lattice will allow optimisation of the bunch length: An isochronous beam optics has been chosen for arc 1,2,3 to allow short bunches. At higher energies, in arc 4,5,6 an efficient emittance control is needed, as the effects of the emitted synchrotron light will take over. These arcs therefore are equipped with a theoretical minimum emittance optics (TME) to mitigate the emittance blow up (see Fig.~\ref{fig:arc_lattice_tme}). 
The magnet structure of the linacs has to provide focusing for the complete energy range of the accelerating / decelerating beams. Here a FoDo structure has been chosen with a phase advance of $130^{\circ}$ per cell. Different cell lengths have been investigated and
simulation studies showed - not unexpectedly - an increasing performance for a shorter cell length.
At the end of the Linac, the beam has to be guided into the return arc that corresponds to the beam rigidity at the given acceleration step.  A combination of dipoles and quadrupole magnets provides the vertical bending and adapts the beam optics to the  arc structure. This  ``spreader" (in front) and ``re-combiner" (after the arc) represent a non-dispersive  deflecting system to provide the necessary vertical off-set between the three arc modules and limit at the same time the detrimental effect on the vertical beam emittance.
\subsection{Interaction Region}
\label{sec:ir}
The Interaction Region (IR) of the LHeC is a most challenging part
of the machine: While seeking for highest luminosity in
$ep(A)$ collisions,
which includes  mini-beta insertions for strong focusing of both beams, the  colliding electron and hadron beams have to be separated after their collisions and guided to their lattice structures, to avoid parasitic bunch encounters. In addition, collisions and beam-beam effects with the second non-colliding proton beam  have to be avoided
and backgrounds such as beam halo be mitigated.
\subsection{Proton Beam Optics}
\begin{figure}[htb]
    \centering
    \includegraphics[width=0.48\textwidth]{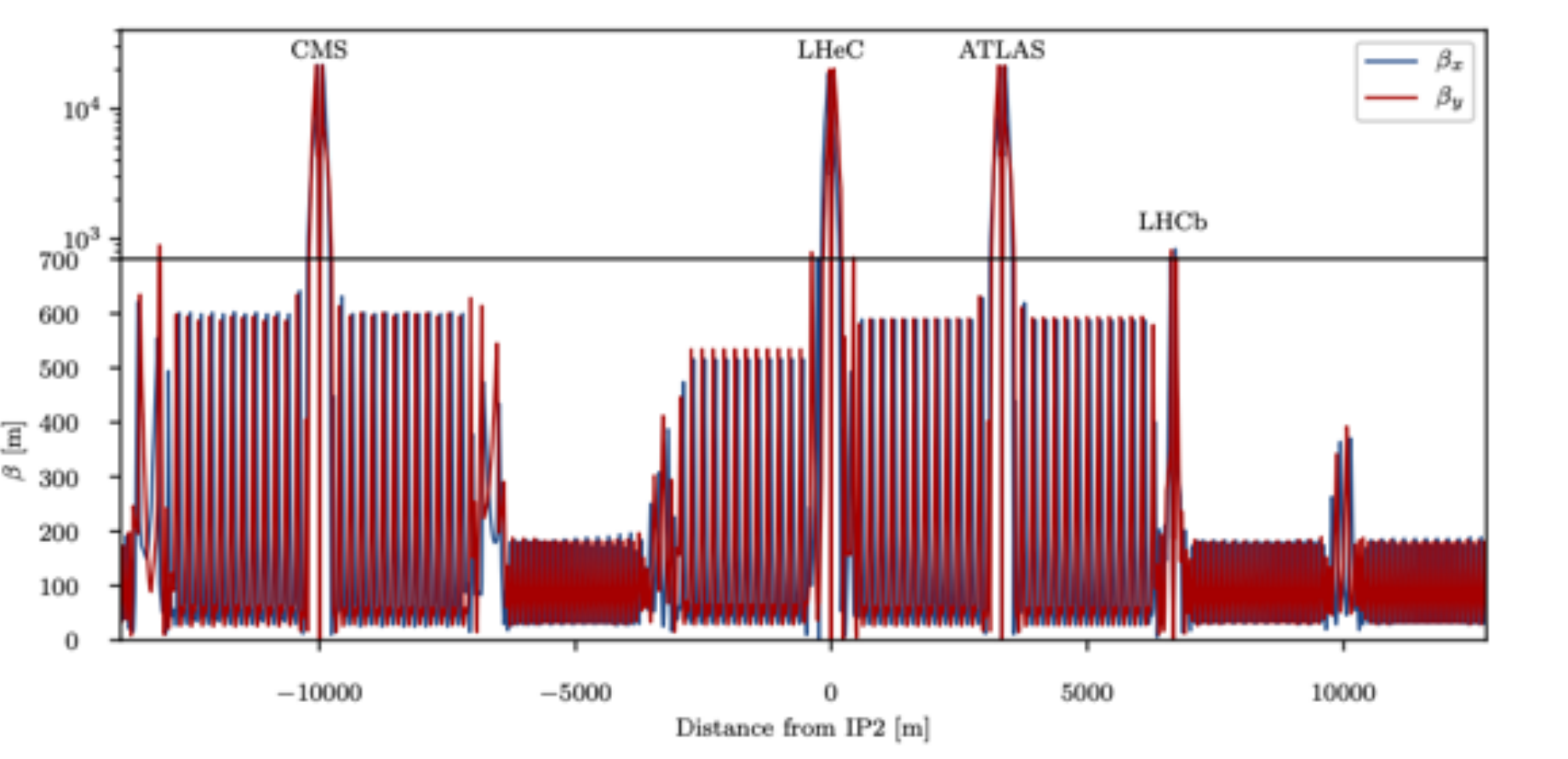}
    \caption{LHC proton beam optics, optimised for the LHeC design values at the LHeC IP.}
    \label{fig:p_optics}
\end{figure}

The optics of the colliding proton beam follows the standard settings of the HL-LHC. Fig.~\ref{fig:p_optics} shows the proton optics  at the interaction point of the LHeC. The long-ranging beta-beat which is an essential feature of the HL-LHC optics~\cite{HL-LHC}  is clearly visible on both sides of the IP and will be used for both, h-h and e-p collisions in IP2.

\begin{figure}[htb]
    \centering
    \includegraphics[width=0.45\textwidth]{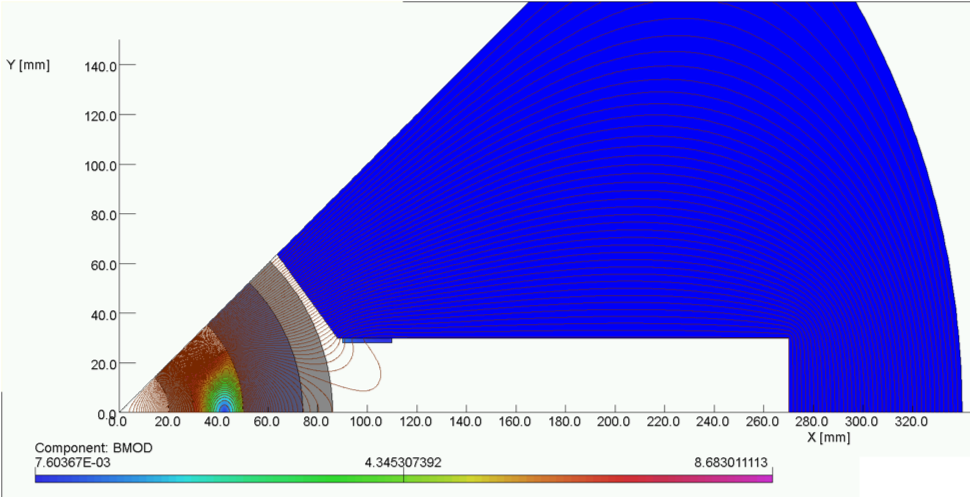}
    \caption{Layout of the first proton quadrupole after beam separation. Special emphasis is put on minimising the remaining field in the electron aperture at a distance of 106\,mm from the p design orbit.}
    \label{fig:p_quad}
\end{figure}

Special design effort is needed in the layout of the super conducting quadrupole “QA1”: Positioned right after the electron mini beta quadruples, it has to provide sufficient aperture and gradient to re-match the proton optics towards the arc structure. At the same time a field free region inside the cryostat is needed for the outgoing electron beam. Figure ~\ref{fig:p_quad} shows a first layout of the magnet. The field calculations for both apertures are determined using the magnet design code ROXIE~\cite{Roxie} with special emphasis on minimizing the remaining quadrupole field in the electron aperture: located at a distance of \SI{106}{mm} from the proton design orbit - it has to be low enough not to distort the electron beam. Following the first layout and field calculations described, further R\&D will be needed leading to a detailed design and construction of a prototype magnet in order to investigate the feasibility of the technical concept.
\subsection{Electron Beam Optics and Separation Scheme}
\label{ElectronBeamOptics}
\begin{figure*}[htb]
    \centering
\includegraphics[width=0.9 \textwidth]{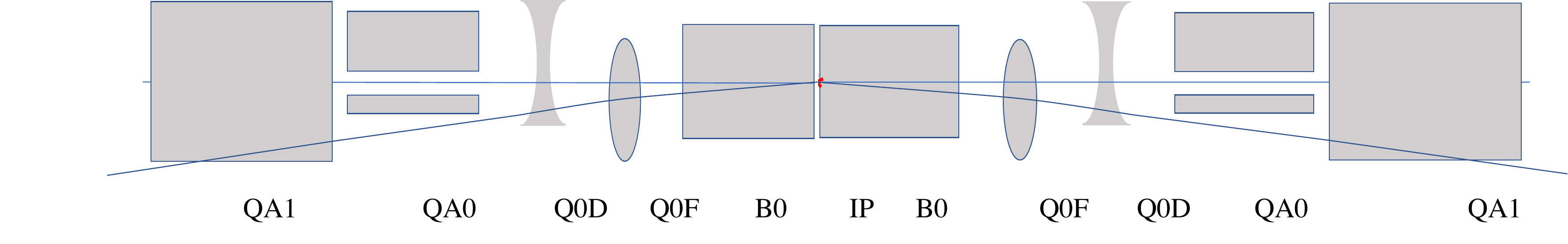}
    \caption{Schematic view of the combined focusing \& beam separation scheme }
    \label{fig:sep_scheme}
\end{figure*}

\begin{figure}[htb]
\includegraphics[width=0.48 \textwidth]{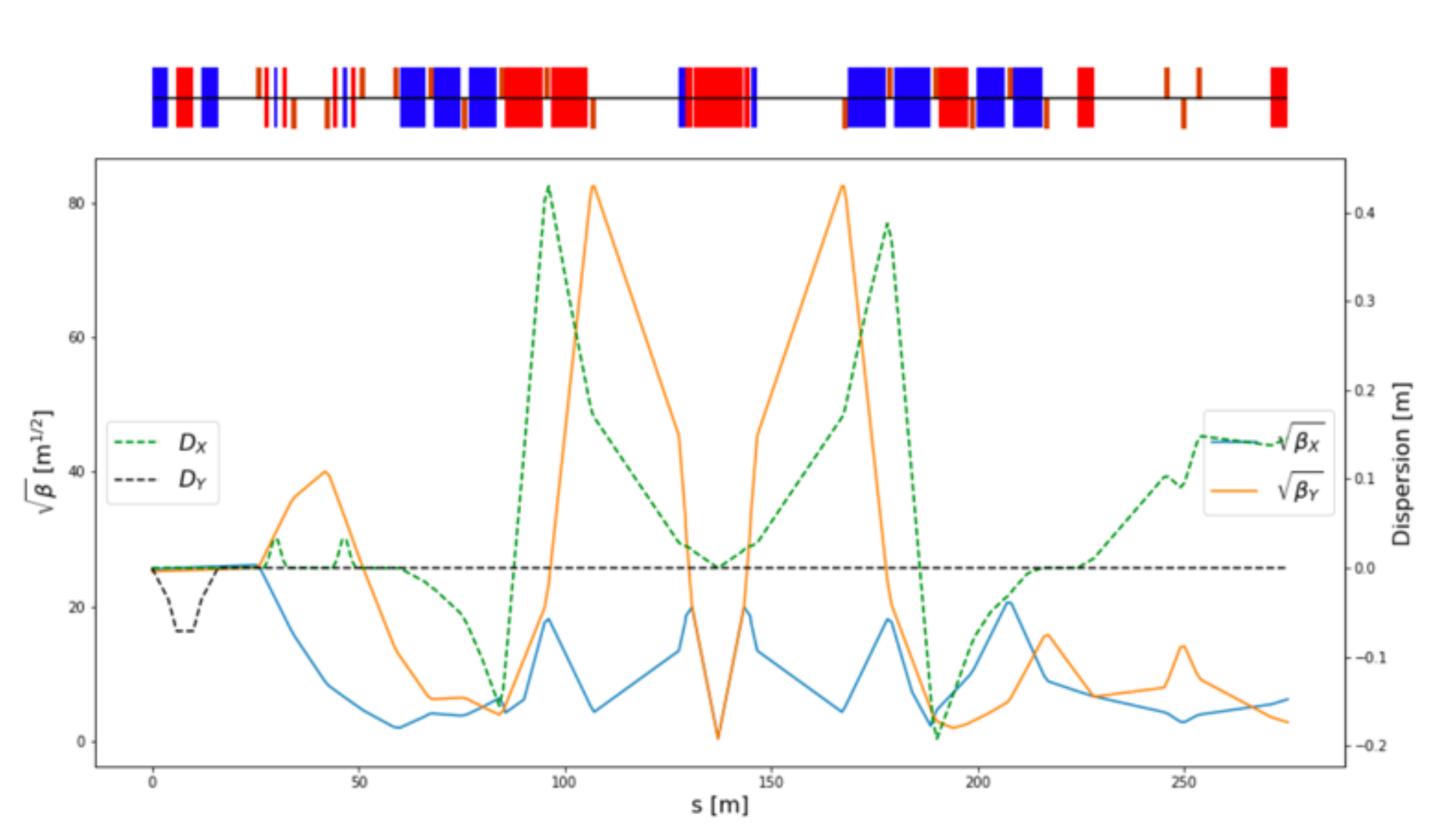}
    \caption{Optical functions of the electron beam in the IR.}
    \label{fig:e_optics}
\end{figure}

The design orbit of the electron beam - accelerated by the ERL and brought into collision at IP2 - will be merged with the proton orbit only in a short part of the lattice: Due to the different beam rigidities, 
$$(B \ast \rho)_p = \SI{23333}{\tesla \meter} \hspace{0.5cm}
(B \ast \rho)_e = \SI{167}{\tesla \meter}$$
a common focusing structure is not possible. 
The design of the IR therefore has to take a manifold of conditions into account:
Focus the electron beam to the required $\beta$ values in both planes, establish sufficient beam separation, optimise for smallest  critical energy and synchrotron light power, and leave sufficient space for the detector hardware. 
A separation scheme has been established~\cite{Kevin} that combines these requirements in one lattice structure (see Fig.~\ref{fig:sep_scheme}). 
Due to the different rigidity of the beams, a separation is possible through the common effect  of several  magnetic fields: The spectrometer dipole of the LHeC detector, named {\it B0} in the figure, is used to establish a first separation. Right after and as close as possible to the IP, the mini-beta quadrupoles of the electron beam are located. They provide focusing in both planes for matched beam sizes of protons and electrons at the IP: 
$$
\beta_x(p) = \beta_x(e), \hspace{0.5cm} \beta_y(p) = \beta_y(e)
$$ 
Moreover, they are positioned off-center with respect to the electron beam, thus acting as combined function magnets to provide the same bending field as the separator dipole: A 
quasi-constant,  soft bending of the electron beam is achieved throughout the magnet structure: $ 1/{\rho_{B0}} =1/ \rho_{quad_f} =1/ \rho_{quad_d}   $. Additional conditions were put for a reduced beam size of the electron beam at the location of the first proton quadrupole. At this position, $L^*$=\SI{15}{m}, the reduced electron beam size leads automatically to a minimum of the required beam separation and as direct consequence to smallest synchrotron radiation effects. 
The optical functions of the electron beam in this optimised interaction region are shown in Fig.~\ref{fig:e_optics}. 

\subsection{Joint $eh$/$hh$ Operation Options}
\label{Cocurrent_Operation}
\begin{figure}[thb]
    \centering
    \includegraphics[width=0.4 \textwidth]{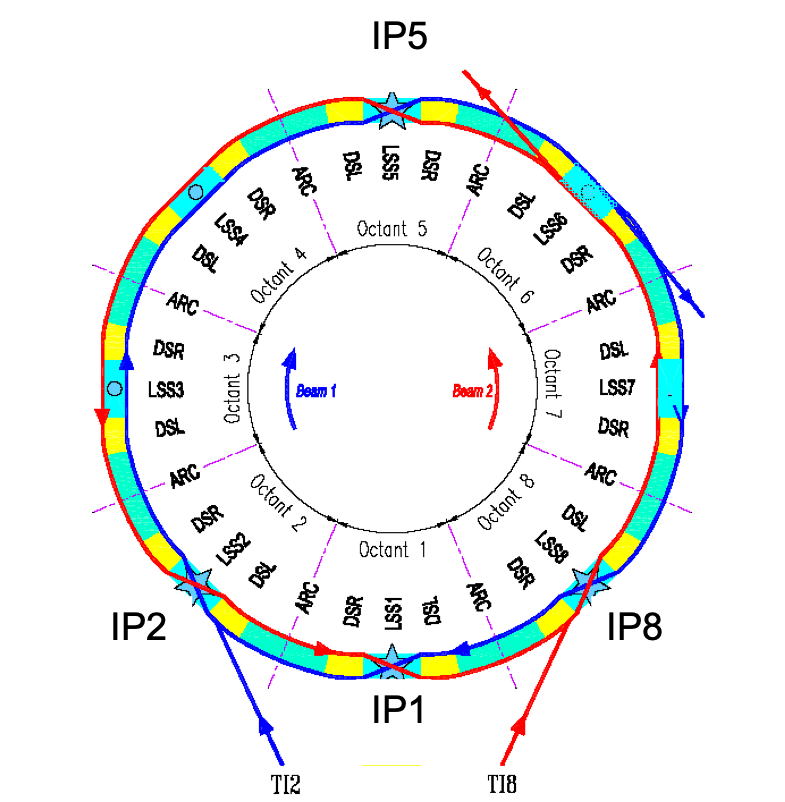}
    \caption{Geometry of the two LHC beams, crossing from inner to outer ring in the four interaction points IP1,2,5 and 8.}
    \label{fig:lhc_geometry}
\end{figure}

The interaction region layout described above has been optimised for highest luminosity, matched beam sizes between electrons and the colliding proton beam and a smooth but efficient beam separation scheme. Still, an additional boundary condition arises from the second, ``non-colliding'' proton beam: A concurrent operation of the LHeC as electron-proton collider means that the operation as $ep$ collider will be possible in parallel to the standard LHC proton-proton operation. During $ep$ operation in IP2, with electrons provided by the ERL, the standard $pp$ collisions in the LHC interaction points IP1, (ATLAS), IP5 (CMS) and IP8 (LHCb) will continue and thus the second proton beam has to be guided through the new interaction region IR2, in parallel to the electron and proton beams. 
At IP2 therefore, in $ep$ operation mode, the second non-colliding proton beam will be separated by a symmetric orbit bump to avoid direct collisions between the two proton beams as well as with the electron beam. Parasitic encounters with the subsequent bunches are suppressed by a vertical crossing angle. This scenario follows the LHC standard operation, where similar orbit bumps are applied during injection and acceleration phase of the two beams.
Additional constraints arise from the need to 
 preserve the overall LHC geometry: The two LHC proton beams will have to cross over from the inner ring to the outer, see Fig.~\ref{fig:lhc_geometry}. 
All in all,  two basic operation modes have to be established:
\begin{itemize}
    \item Standard $hh$ collisions in IP~1,2,5,8, no electron beam.
    \item Concurrent operation of $eh$ collisions in IP2 and  $hh$ collisions in IP 1,5,8. 
\end{itemize}

Concerning the first operation mode, the set up will be equivalent to the HL-LHC upgrade lattice and optics, with the two hadron beams colliding in all LHC interaction points. The magnets of the electron mini beta structure and beam separation scheme, shown schematically in  Fig.~\ref{fig:sep_scheme}, will be switched off.

\begin{figure}[thb]
    \centering
    \includegraphics[width=0.45 \textwidth]{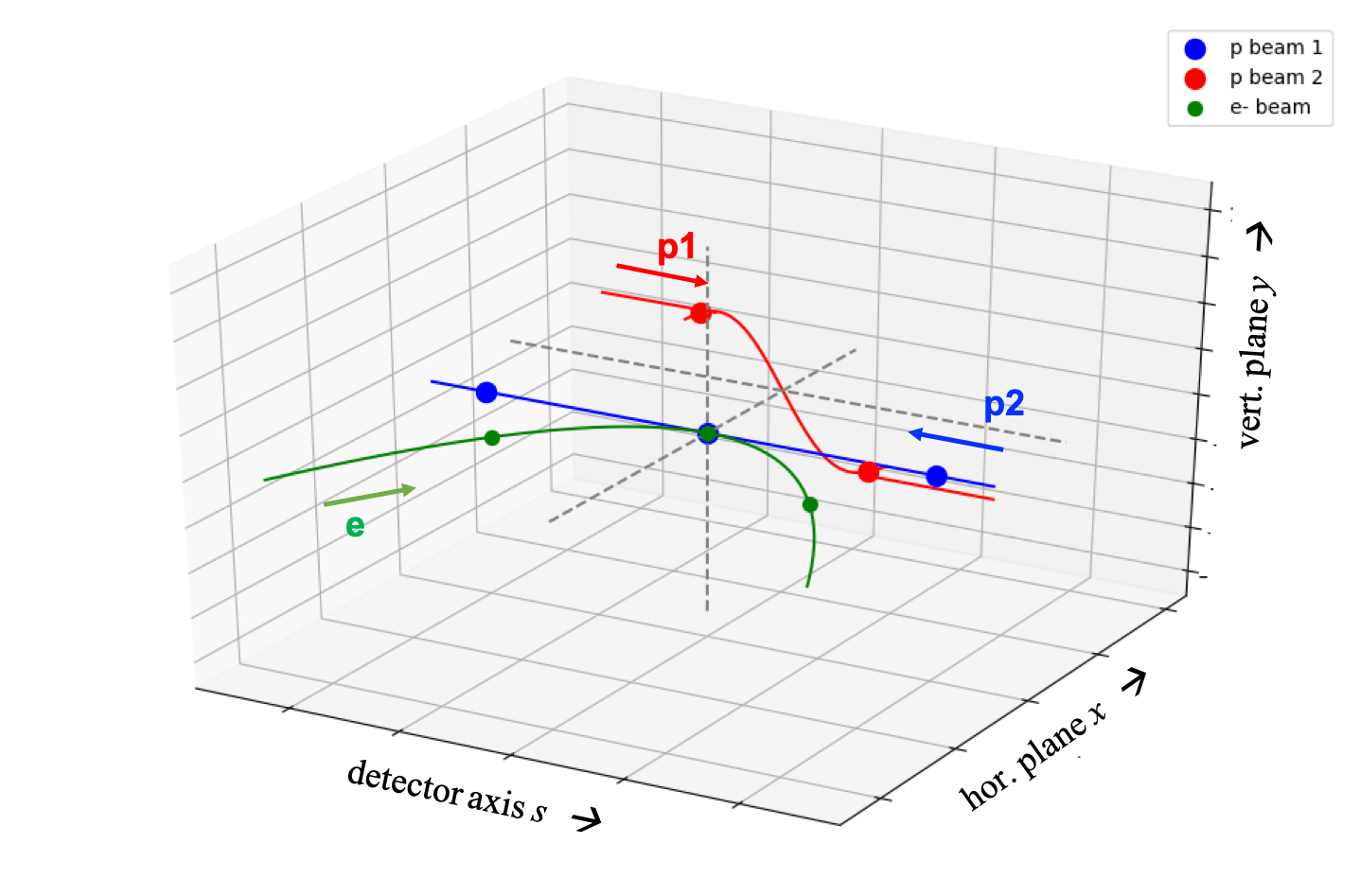}
    \caption{Schematic view of the three beams in the interaction region. Collisions between electrons and proton beam 1 and a well separated proton beam 2.}
    \label{fig:p-ep_scheme}
\end{figure}

\begin{figure}[htb]
    \centering
    \includegraphics[width=0.45 \textwidth]{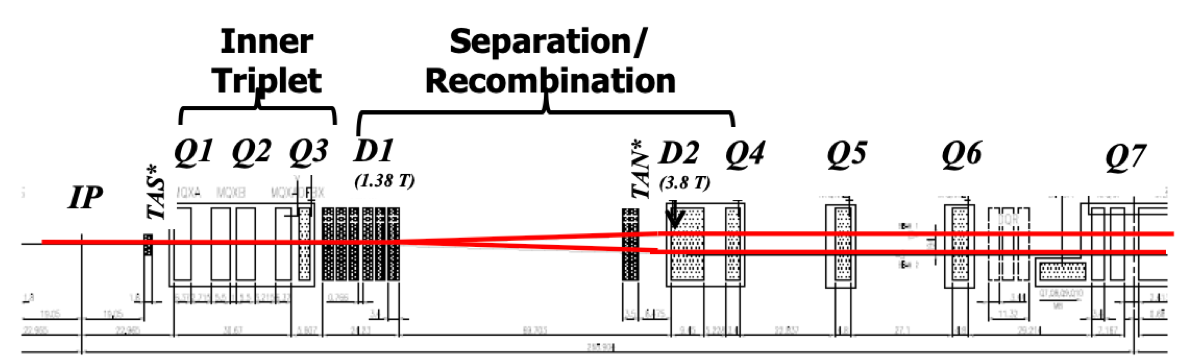}
    \caption{Schematic view of the LHC proton beam separation scheme. The two separator dipoles D1 and D2 provide the hor. separation needed, before the beams enter their distinct magnet lattices in the arcs.}
    \label{fig:LHC_sep_scheme}
\end{figure}

For the second operation mode the colliding hadron beam (protons or ions)  will be focused to match the size and position of the ERL electron beam at the IP. Electron beam focusing and beam separation between electrons and hadrons will follow the above mentioned scheme and due to the flexibility of the ERL optics a wide range of parameters (i.e. beam sizes) at the IP can be achieved, fulfilling the requests of both, e-p and e-A collisions. The second non-colliding LHC beam, however, will pass untouched through IR2, but still being used for collisions and data taking in IP~1,5 and 8.  
For this purpose, a sufficient beam separation between this non-colliding proton and the colliding beams in IP~2 is needed. 
Schematically the situation is shown in Fig.~\ref{fig:p-ep_scheme}.
The beam separation is established via the LHC standard separation bumps, that are used during beam injection and throughout the complete acceleration phase. 
At the interaction point IP2, direct collisions are avoided by a horizontal offset of the non-colliding beam. In addition a vertical crossing angle is applied to prevent effects from the so-called parasitic encounters, that otherwise would occur at a distance of  half a bunch spacing  (\SI{25/2}{ns}). While this scheme is used during LHC standard operation it requires special attention for the concurrent $ep$/$pp$ operation. A special beam optics for the non-colliding proton beam has to be established, to provide sufficient aperture for this new type of beam operation. The colliding proton beam will be focused strongly to achieve a $\beta$-function of \SI{10}{cm} at the IP. At the same time the non-colliding beam will see a relaxed optics with smallest achievable beam size in the proton mini beta quadrupoles. First estimates, based on an ``injection type optics''   with $\beta^*$ of \SI{15}{m} lead to an additional aperture request of about 10\% in the first proton quadrupole Q1A. Further downstream the two proton beams will follow the usual beam separation defined by the separator dipoles D1 and D2 (see Fig.~\ref{fig:LHC_sep_scheme}). Further studies will concentrate on the level of flexibility of the different LHC magnet lattices which is a pre-requisite for the proposed scenario. Beyond that, the beam-beam effect between the electron and the non-colliding proton beam -- traveling for a considerable distance in parallel to each other -- will be studied in detail.

\subsection{Synchrotron Light}
The synchrotron light parameters, i.e.\ critical energy, radiation power and the geometry of the emitted light cone   were determined with the simulation code BDSIM~\cite{bdsim}. As expected, the synchrotron light conditions in the arcs become more serious turn by turn, reaching the highest level in the return arc 6, after the collision point. The values are summarised in Tab.~\ref{tab:sy_li_arc}. Special care is needed in the vicinity of the particle detector. The properties of the focusing elements, the separation scheme and the geometry of the  interaction region have been optimised for  smallest critical energies and power of the emitted light.
\begin{table}[htb]
    \centering
    \begin{tabular}{|c|r|r|r|}
    \hline
    Arc & Energy & Crit. Energy &  Power \\
        &   (GeV)   &   (keV)    &    (MW) \\
    \hline 
    1 & 8.75    &  3.2     & 0.01  \\
    2 & 17.00   &  23.9    & 0.21  \\
    3 & 25.25   &  78.5    & 0.75  \\
    4 & 33.5    &  183.3   & 2.45  \\
    5 & 41.75   &  354.8   & 5.87  \\
    6 & 50.0    &  609.3   & 12.17 \\
    \hline
    \end{tabular}
    \caption{Critical energy and power of the emitted synchrotron light in the return arcs of the ERL.}
    \label{tab:sy_li_arc}
\end{table}

\begin{figure}[htb]
    \centering
    \includegraphics[width=0.4\textwidth] {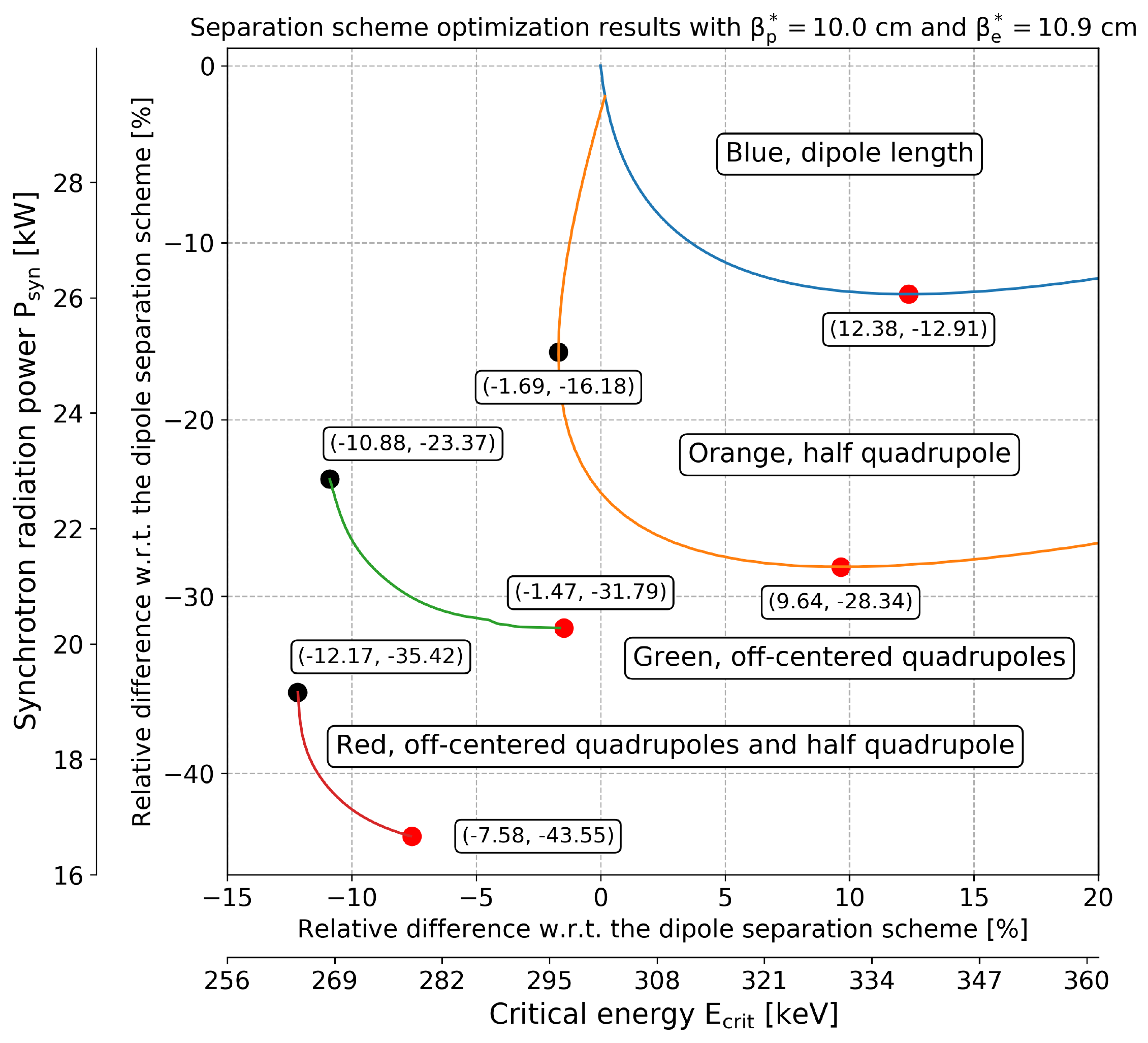}
    \caption{Optimising the synchrotron light for lowest critical energy and power in the IR, details in the text.}
    \label{fig:sy_li_opti}
\end{figure}

 Fig.~\ref{fig:sy_li_opti} summarizes the results. The graph shows the reduction of the critical energy and power in the interaction region, due to the different steps of the optimisation procedure. Starting from a pure separator dipole design to establish the required beam separation, the concept of a  half-quadrupole as first focusing element in the proton lattice is introduced as well as an improved beam separation of the electrons by off-centre quadrupoles.
The actual distribution of the detector dipole field and the off-centre quadrupoles has a considerable effect: The red and black points in the graph correspond to the minimum achievable critical energy and emitted power, respectively.  Dedicated calculation of the synchrotron light cone and a sophisticated machine detector interface including absorbers will be needed to shield the detector parts and accelerator magnets. 

\subsection{Beam-Beam Effects}
The beam-beam effect will always be the final limitation of a particle collider and care has to be taken, to preserve the beam quality and to limit  detrimental effects on the emittance for assuring a successful energy recovery process in the ERL. 

\begin{figure}[!tbh]
    \centering
    \includegraphics[width=.8\columnwidth]{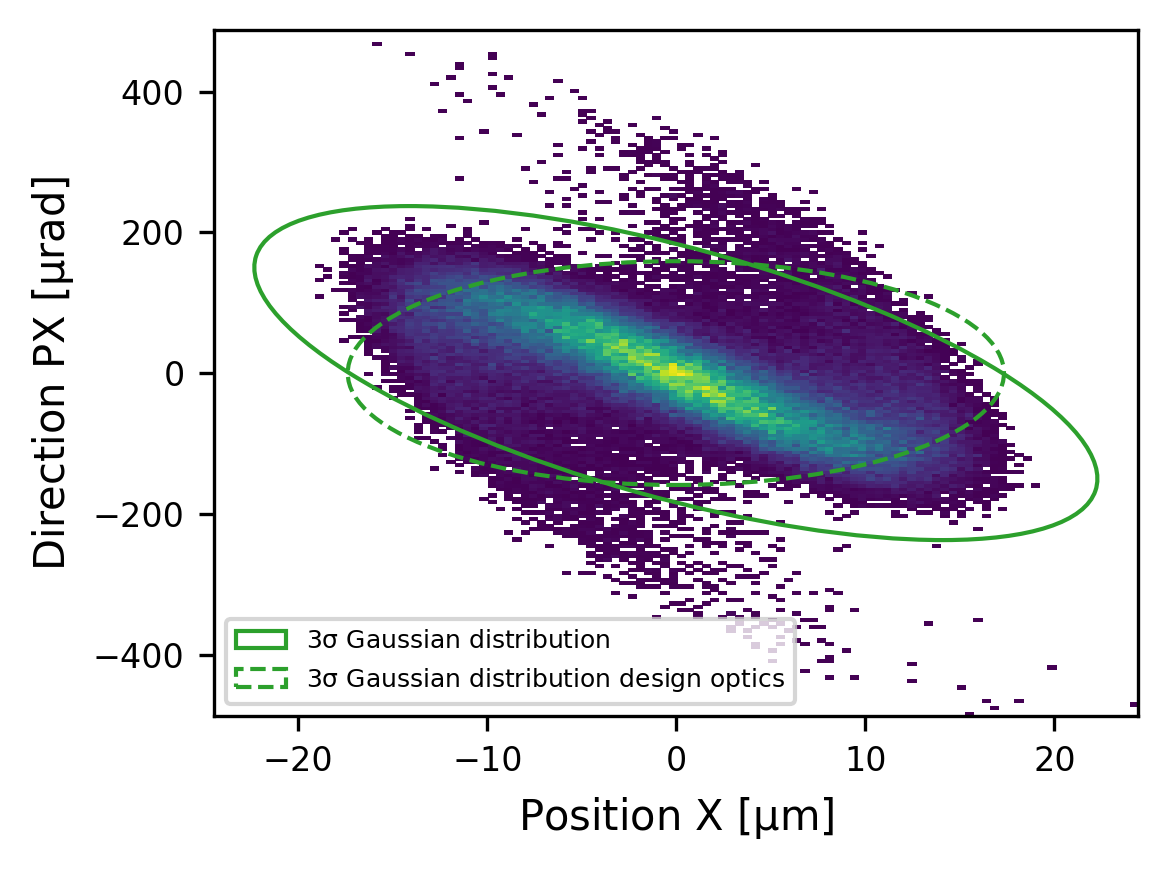}
    \caption{\label{fig:BB_phasespace}Phase space of the electron distribution after beam collision, backtracked to the IP for matched optics conditions of electrons and the HL-LHC proton beam.}
    \label{fig:electronphasespace}
\end{figure}

The beam-beam interaction has been simulated with a weak strong tracking simulation for a matched transverse beam size of the electron  and proton beam at the IP. In Fig.~\ref{fig:BB_phasespace} the situation post collision is represented in the (x,x') phase space.
While tails in the transverse beam distribution as consequence of the beam-beam effect are clearly visible, the core of the beam  still remains  in a quasi ellipse like boundary. The coordinates obtained are used as starting conditions for the deceleration part of the ERL for a full front-to-end simulation.

The resulting emittance increase and luminosity, taking into account the beam-beam force are summarised in Tab.~\ref{tab:bb_lumi}.
\begin{table}[htb!]
  \centering
  \begin{tabular}{l|c}
  Parameters & Optical matching \tabularnewline 
  \hline 
  Luminosity  & \SI{8.2e33}{\per \square \centi \meter \per \second}\tabularnewline
  $\Delta \gamma\varepsilon$  & \SI{15}{\milli \meter \milli \radian} \tabularnewline
  \hline
  \end{tabular}
  \caption{\label{tab:optimumluminosity} Luminosity and transverse emittance growth for the optical matching.}
     \label{tab:bb_lumi}
\end{table}
The beam-beam effect on the proton bunch remains in the shadow of the other effects and is considered as not critical. A careful alignment of the electron bunch at the IP, however, will be necessary as it could lead to undesirable proton emittance growth build up~\cite{Agostini:2020fmq}.

The phase space distribution of the electrons after beam collision does not follow a Gaussian distribution. The non-linearity of the interaction distorts the electrons on the edges as well as modifies the Twiss parameters from the original design as shown in Fig.~\ref{fig:electronphasespace}. The distortion of the phase space impacts the particle density and makes the core and tail of the distribution more populated than a Gaussian distribution. Nevertheless, the ellipse fitted to the post-collision distribution, that takes into account the modification of the Twiss parameters at the interaction point -  including a so called capture optics - has a higher central density and the tails are  less populated.

\begin{table}[htb!]
  \centering
  \begin{tabular}{lcc}
   & Gaussian distribution & Optical matching \tabularnewline  \hline
  1 $\sigma$ & 68.3\% &  46.3\% (70.7\%) \tabularnewline
  2 $\sigma$ & 95.4\% & 78.4\% (95.4\%) \tabularnewline
  3 $\sigma$ & 99.7\% & 95.8\% (98.4\%) \tabularnewline
  4 $\sigma$ & 99.9\% & 99.2\% (99.5\%) \tabularnewline \hline
  \end{tabular}
  \caption{Comparison of the electron distribution after non linear beam-beam interaction. The values represent the density of electrons for several rms emittance areas for the design optics as well as for the fitted ellipse of the post-collision distribution, in parenthesis.}
\end{table}

Further studies are needed regarding the impact of a smaller beam size of the electrons at the IP \textit{e.g.} following the quest for a luminosity optimum. In fact, the optimal separation scheme may need to be adapted, the beam stay-clear aperture in the mini-beta quadrupoles would decrease and could be a showstopper for this luminosity optimisation scheme and finally the use of not matched lepton/hadron beam sizes could lead to instability for the proton bunch.
 
\subsection{Front-to-End Tracking Studies}

The tracking simulations of the ERL have been performed with the tracking code PLACET2~\cite{PLACET2} and include, beyond the properties of the magnetic fields,  the Incoherent Synchrotron Radiation (ISR) and the weak-strong beam-beam interaction at the interaction point. The studies focused on the achieved transmission and the beam quality along the ERL passages, \textit{i.e.} the emittance budget required, for different machine circumferences that are considered for the basic machine layout. The beam parameters used for the tracking simulations correspond to the main parameter list, see Tab.~\ref{tab:main_parameters}.

\begin{figure}[!tbh]
    \centering
    \includegraphics*[width=.95\columnwidth]{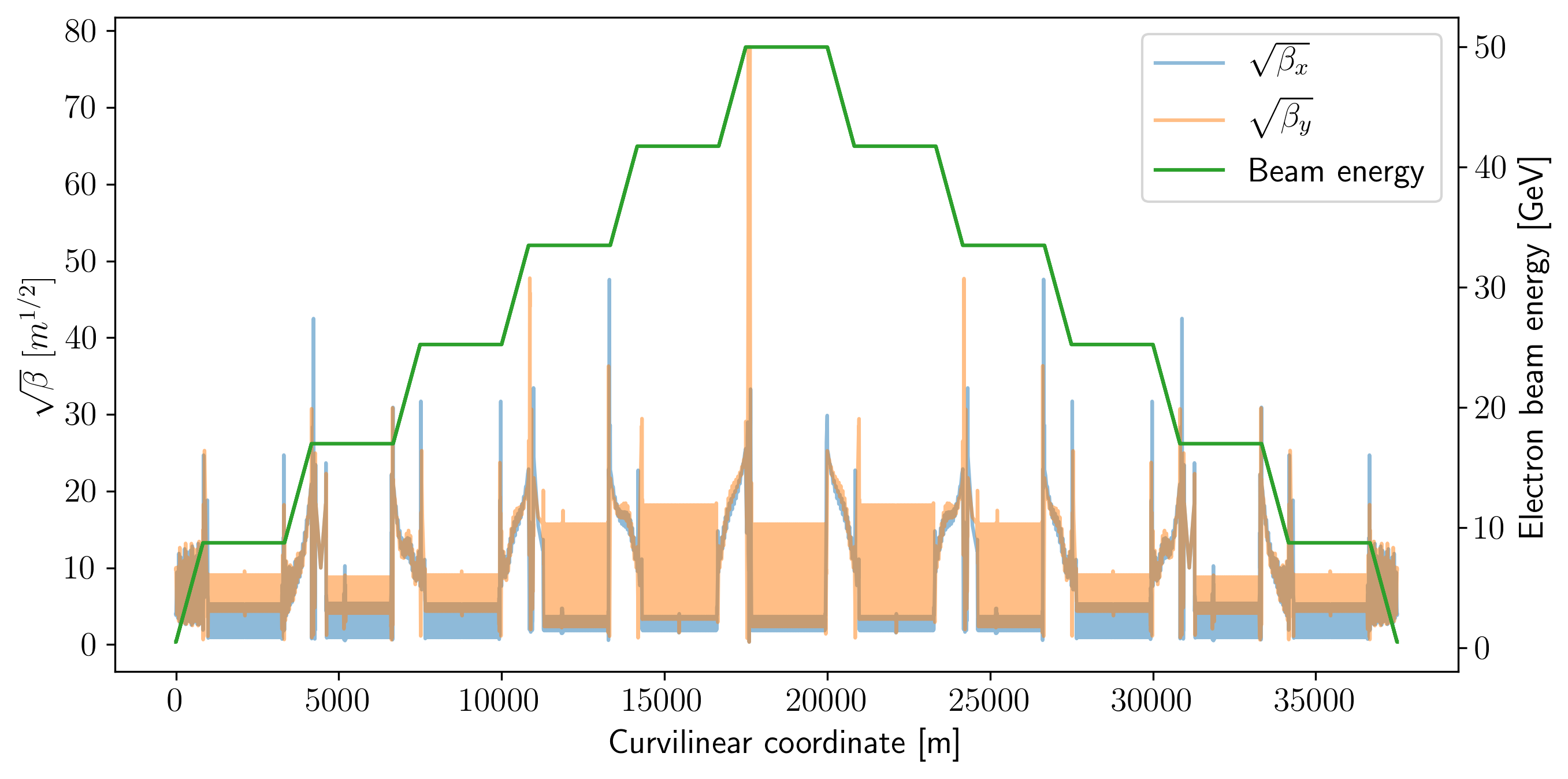}
    \caption{\label{fig:multiturnoptics}Representation of the beta functions and the beam energy along the multi-turn ERL operation.}
\end{figure}

The optics design of the multi turn ERL is shown in Fig.~\ref{fig:multiturnoptics} and presents the sequence of linacs and arcs leading to the interaction region with a strong focusing and accordingly large vertical beta function in the mini beta quadrupoles. The other peaks are located in the matching sections between the linac optics and the periodic arc structure.
The tracking takes place over three acceleration turns until the IP. Three deceleration turns are following in the same lattice structure, established via a RF phase shift in the highest energy return arc 6.

The objectives are: obtain the required transverse emittance at the IP; collide with the proton beam;  minimise the emittance growth; taking into account eventual optics mismatch and  distortion due to the non-linear beam-beam effect; decelerate the electron beam during the energy recovery process and guarantee minimum particle losses, while the energy spread will reach levels of a few percent in the last deceleration step.

The synchrotron radiation for each ERL circumference that has been studied varies significantly and has a strong influence on the energy recovery efficiency, see the results Tab.~\ref{tab:results}. 

\begin{figure}[!htb]
    \centering
    \includegraphics[width=.95\columnwidth]{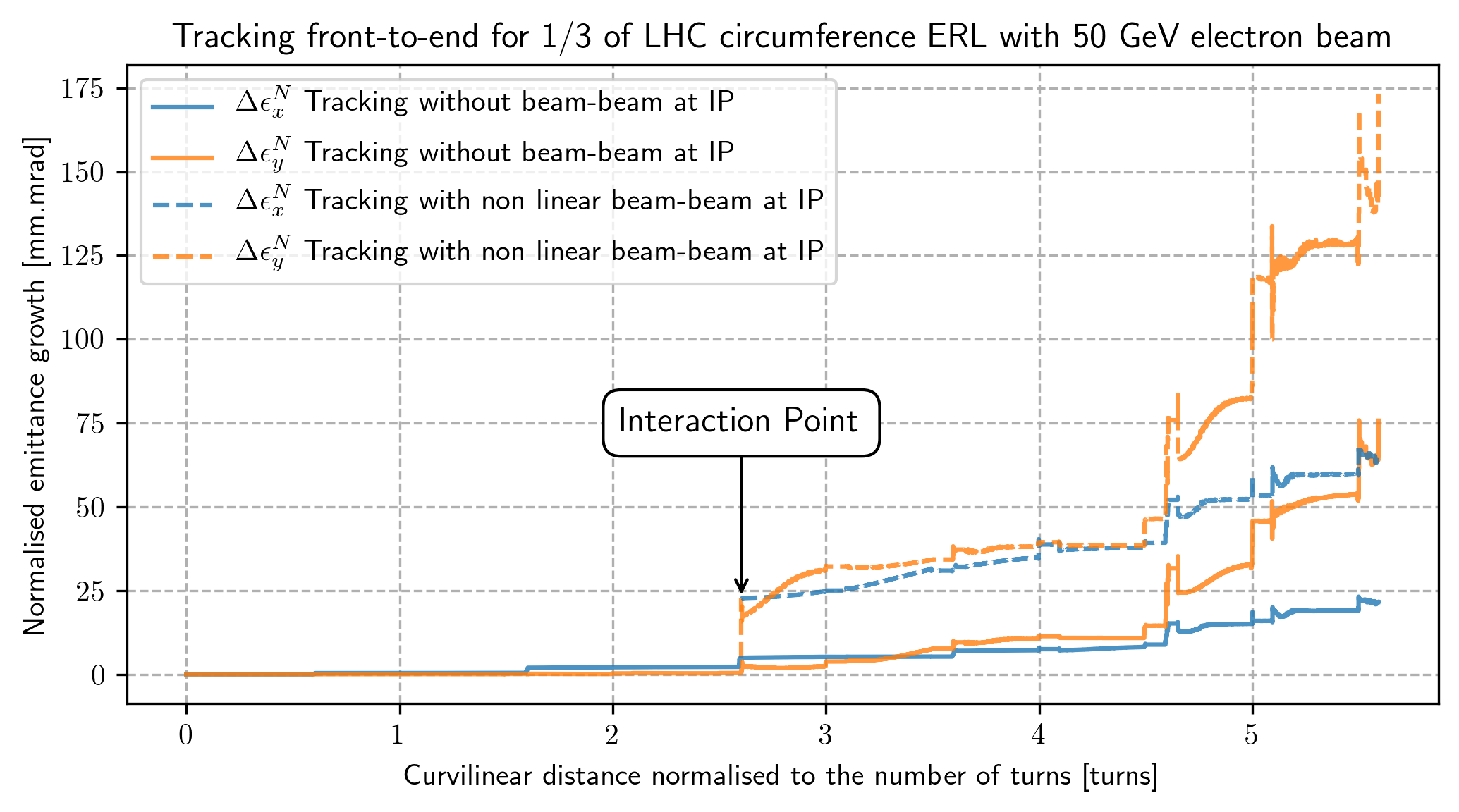}
    \caption{\label{fig:emit_growth} Emittance growth along the curvilinear coordinate for the largest ERL design, corresponding to  1/3 of the LHC circumference.}
\end{figure}

The results of the tracking simulations and the obtained emittance growth during the three turn beam acceleration agree nicely with the analytical calculations. After the interaction region the particles increasingly gain energy spread that creates a deviation from the design optics. The optics mismatch results in an extra emittance growth and ultimately leads to beam losses during the deceleration phase. The results of the emittance growth  for the largest LHeC circumference studied, 1/3 of LHC, can be found in Fig.~\ref{fig:emit_growth}.

The recent tracking studies demonstrated that the electron beam quality can be preserved until the IP in order to meet a normalised transverse emittance of \SI{30}{\milli \meter \milli \radian} at the interaction point. Then, it is followed by a strong non-linear beam-beam interaction and finally decelerated over 3 turns to be dumped at \SI{500}{\mega \electronvolt}. The tracking results of all the ERL circumferences studied, also including the synchrotron radiation and the beam-beam disruption, give an excellent transmission of close to 100\%, see Tab.~\ref{tab:results}. The energy recovery efficiency is mainly constrained by the synchrotron light losses in the arcs that need to be compensated by extra RF cavities. It can be noted that the  ERL designs that consider smaller machine circumferences require a smaller horizontal injection emittance that will potentially not allow enough margin for further studies including magnet field errors and misalignment.

\begin{table}[tbh!]
    \centering
    \small
    \begin{tabular}{l|ccc}
         ERL size  & 1/3 $C_{LHC}$& 1/4 $C_{LHC}$ & 1/5 $C_{LHC}$\\
         \hline
         $\gamma \varepsilon_x^{\text{inj}}$ [\si{\micro \meter \radian}] & 25.4 & 22.7 & 15.1 \\
         $\Delta p /p$ at IP  & 0.021 \% & 0.029 \% & 0.041 \% \\
         transmission & 99.9 \% & 98.9 \% & 98.4 \% \\
         energy recovery & 97.9 \% & 96.7 \% & 95.4 \%  \\ \hline
    \end{tabular}
    \caption{\label{tab:results}Results of the tracking simulations including beam-beam effect and synchrotron radiation for different ERL designs.}
\end{table}

\section{Joint HI Physics and a Detector} 
\label{jointehhh}
The initial studies on a common  IR for electron-hadron and hadron-hadron scattering, as presented above, suggest to indeed think of its possible joint use. Subsequently, 
a heavy-ion physics programme is sketched which would be enabled by combining $eh$
and $hh$ opportunities. The section also presents a sketch of how the LHeC detector
would need to be modified when used also for $hh$ collisions. Would this promising
direction be followed, a joint design concept would be a next natural step, regarding obviously the beam pipe, particle ID, hadronic calorimetry, end cap and other design features.

\subsection{Heavy Ion Physics}

The physics opportunities provided by the availability in the same detector of DIS off nuclei, proton-nucleus and nucleus-nucleus collisions, $eA$, $pA$ and $AA$ respectively, 
are immense (see e.g. the discussions in~\cite{Strategy:2019vxc} and refs. therein):
\begin{itemize}
\item On the one hand, as extensively discussed in~\cite{AbelleiraFernandez:2012cc,Agostini:2020fmq}, $eA$ collisions at high energies at the LHeC will reveal the partonic structure of nuclei and the QCD dynamics in hitherto unexplored kinematic regions of high energies and parton densities. This is the region of relevance for $pA$ and $AA$ collisions at the LHC and beyond.
\item On the other hand, the proposed heavy-ion (HI) detector to be installed in IP2 during LS4 to operate during subsequent LHC Runs~\cite{Adamova:2019vkf} aims to provide outstanding tracking capabilities in the soft region down to tens of MeV and, due to fast timing, large possibilities for PID beyond $dE/dx$~\cite{Preghenella:2020mxn}, and to be able to work and record minimum bias collisions at the largest AA achievable luminosity.
\item The combination of outstanding track reconstruction, extended to $\sim 1$\,degree in the backward and forward directions, and particle ID in the soft sector, with electromagnetic and hadronic calorimetry and muon detection makes this a general purpose detector for $pp$, $pA$ and $AA$ collisions, with larger capabilities for QCD than ATLAS and CMS and larger acceptance than LHCb.
\end{itemize}
Such detector configuration will be suited to explore the new possibilities for physics with ions after LS4 discussed in~\cite{Citron:2018lsq} as, for example, the larger luminosities provided by ions lighter than Pb (O, Ar, Kr) to analyse the presently least understood stage of hadronic collisions, the initial one~\cite{Apolinario:2017sob,Huss:2020dwe}, by using hard probes. In the following we elaborate on such possibilities.

\subsubsection{Nuclear Structure}

\begin{figure}[htb]
\begin{center}
  \includegraphics[width=0.48\textwidth]{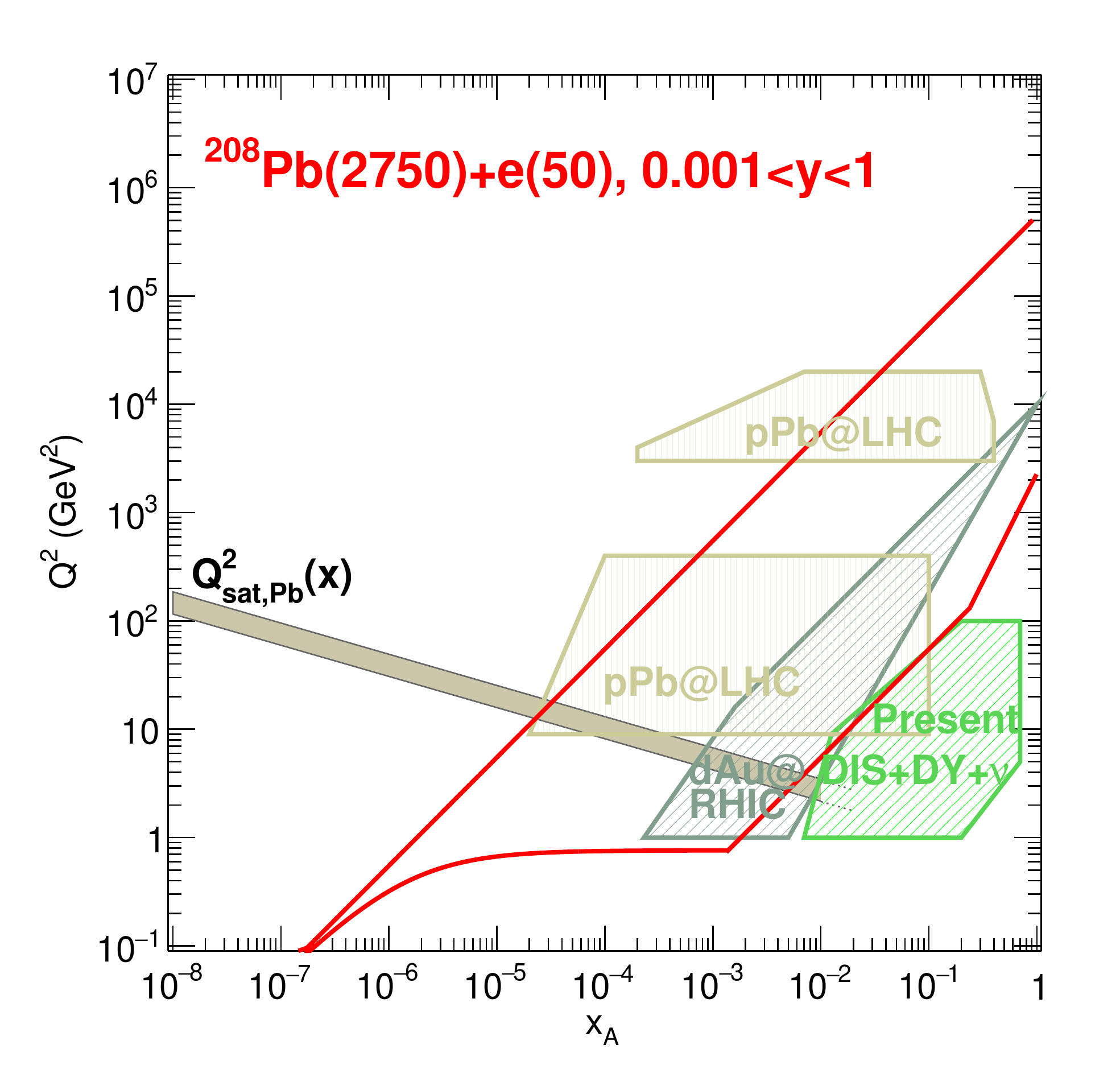}
  \end{center}
\caption{Kinematic plane studied in $e$Pb collisions at the LHeC~\cite{Agostini:2020fmq} (solid red lines) together with the regions explored in present analyses~\cite{Eskola:2016oht}: DIS and DY fixed target data (hatched area in green), hadron production in dAu collisions at RHIC (hatched area in grey) and Run 1 dijet and EW boson studies in $p$Pb collisions at the LHC (hatched upper region in brown). Also shown in the hatched upper region in brown are the expected coverage from  dijets in Run 2~\cite{Eskola:2019dui} and from EW bosons in future Runs~\cite{Citron:2018lsq}, and in the hatched lower region in brown the expectations from Run 2 D-meson analyses~\cite{Eskola:2019bgf} and from DY and photon studies in future LHC Runs~\cite{Citron:2018lsq,ALICECollaboration:2020rog}. LHeC will not only access the saturation region at small $x$, ``below" the $Q^2_{sat}$ line and still small $\alpha_s$, but covering 5 orders of magnitude in $Q^2$ can uniquely unravel the parton dynamics in nuclei.}
\label{fig:kinions}       
\end{figure}

\noindent The kinematic $x-Q^2$ range\footnote{Note that in DIS $Q^2$ denotes the squared virtuality of the exchanged photon, while in hadronic collisions it designates the kinematic variable (transverse momentum, energy, mass,$\dots$) squared which enters as factorisation scale in the PDFs.} to be explored in $eA$ collisions at the LHeC and during future $pA$ Runs at the LHC is shown in Fig.~\ref{fig:kinions} and compared with that of the set of data used in present analyses of collinear nuclear parton densities. LHC data will cover most of the kinematic region also covered by the LHeC (note also that the region between the lower and upper hatched regions in brown can be analysed by DY studies at LHCb), but the extraction of nuclear parton densities in $pA$ and $AA$ collisions relies on the validity of collinear factorisation down to rather low values of $x$ and transverse momenta where other dynamics beyond leading twist perturbative factorisation could be at work. These new dynamics are strongly suggested by the finding at the LHC that many observables in $pp$ and $pA$ behave in a similar manner to that in $AA$, where they are interpreted as signatures of the existence of the Quark-Gluon Plasma (QGP) -- the {\it small system problem}, see~\cite{Nagle:2018nvi} and refs. therein, and \cite{ZEUS:2019jya,ZEUS:2021qzg,h1prelim} for recent studies in DIS at HERA. Besides, even assuming collinear factorisation to hold, the sensitivity to different flavours varies strongly when moving in the kinematic plane. 

DIS offers fully constrained kinematics through the reconstruction of the electron angle and energy, a cleaner theoretical environment where factorisations can be pro\-ven~\cite{Collins:2011zzd} and perturbative calculations and resummations can be pushed to very high orders, and the possibility of full flavour decomposition through the combination of NC and CC and heavy flavour tagging. As shown in~\cite{AbelleiraFernandez:2012cc,Agostini:2020fmq}, these opportunities will be fully exploited at the LHeC, where the nuclear PDFs can be determined with unprecedented precision without requiring prior knowledge of proton PDFs. This leads to a complete change of 
nuclear PDF physics, beyond the huge kinematic extension, as flavour by flavour and for
the gluon one will determine the famous R ratios as a function of $x$ and $Q^2$ which 
will enable the necessary 
distinction of nuclear effects, such as anti-shadowing~\cite{Brodsky:1989qz} for example, 
from the appearance of non-linear parton dynamics.

Factorisation schemes exist beyond collinear factorisation, such as high-energy factorisation, TMD,$\dots$~\cite{Collins:2011zzd}, and, eventually, the breaking of linear evolution when parton densities become large enough with decreasing $x$ or increasing  mass number of the colliding objects. Our current understanding of non-linear QCD dynamics views them as density effects, making both $ep$ and $eA$ equally important inputs to check such explanation. The combination of inclusive, diffractive and exclusive (vector mesons and photons) studies at the LHeC~\cite{AbelleiraFernandez:2012cc,Agostini:2020fmq} will establish the correct factorisation and dynamics in the different kinematic regions. Then, with the relevant non-perturbative information (PDFs, GPDs, TMDs,$\dots$) available, the validity of the corresponding factorisation will be checked in $pA$~\cite{Citron:2018lsq}, thus elucidating the mechanism of particle production in high-energy nuclear collisions.

Finally, the possibility of accelerating ions lighter than Pb will clarify the dependence of  parton densities on the mass number. Therefore, it will eliminate the need of interpolations, based on assumed factorisation of the mass number dependence, between different nuclear species in global fits. This will greatly reduce the theoretical uncertainties inherent to the interpolation procedure.

\subsubsection{Soft Physics}

Soft heavy-ion physics deserves attention for a number of reasons. 
The low transverse-momentum spectrum of di-leptons may be sensitive to the 
restoration of chiral symmetry at high temperatures. Low energy photon spectra
may be influenced by dynamics at the initial state. Physics studied beyond LS4 
would focus on aspects that may not, or not fully, be answered 
before~\cite{Kalweit_ALICE3WS}: concerning the existence of charm-nuclei, hadronisation and thermalisation of beauty via its production and flow, production of multi-charm baryons, temperature evolution of the fireball via transverse momentum differential di-lepton measurements or $\rho-a_1$ mixing.

While the standard description of these collective features~\cite{Nagle:2018nvi} is done in the framework of relativistic hydrodynamics, and the comparison with data used to extract QGP properties, it is known that hydrodynamics works well in out-of-equilibrium situations. In fact, it is currently believed that hydrodynamics is the long wavelength limit of quantum field theories. How this macroscopic description emerges from the microscopic QCD dynamics off the highly out-of-equilibrium initial conditions is the hottest topic in the field. To clarify this, it is crucial to establish the proper factorisation at work in $pp$, $pA$ and $AA$ collisions at high energies and the dynamics in the initial stages prior to the application of hydrodynamics. It is here where the contribution from $ep$ and $eA$ collisions in similar kinematic regions -- at the LHeC -- becomes key, as DIS is the ideal system to elucidate these aspects. It will also contribute to reducing the uncertainties in the extraction of QGP properties from the comparison of data with hydrodynamic calculations, those regarding the initial conditions and the initial stage dynamics~\cite{Heinz:2013th}. Additional information on the initial stages of the collision and the thermalisation problem can be accessed using hard probes that demand high transverse momentum particle detection and calorimetry.

\subsubsection{Hard Probes: Vector Mesons and Jets}

\noindent Among the hard probes, heavy-quarkonium production processes have always been a subject of special interest in high-energy physics. They involve both perturbative and nonperturbative aspects of QCD, corresponding to the production of the heavy-quark pair and its non-perturbative evolution. In addition to the hadronic experiments, the LHeC is particularly suited to study  electro- and photo-production of quarkonia up to 
high energies~\cite{AbelleiraFernandez:2012cc}. These processes, which involve a highly virtual photon for electro-production or a real one for photo-production, provide unique opportunities for the study of the quarkonium production mechanism and the perturbative QCD calculation reliability. Moreover, the high gluon densities involved in these processes offer the opportunity to have an insight into the gluon generalised parton distribution in nuclei, the role of color correlations, and the color-dipole nature of quarkonia in unexplored regions far beyond those at HERA and the EIC.  

At low transverse momentum, the proposed HI detector~\cite{Adamova:2019vkf}  at IP2 will offer the possibility to study separately the prompt and non-prompt quarkonium production with the identification of the contribution from excited states by detecting low energy photons. Such separation will allow a better characterisation of the QGP~\cite{Andronic:2015wma}, based up to now on the anomalous nuclear dependence of quarkonium hadroproduction.  Besides, the capabilities of the detector can have a great impact on the field of hadron spectroscopy, opening the possibility to measure the photoproduction of X, Y, Z states. Such studies demand an understanding of the production mechanism of quarkonia which presents large uncertainties until now, and of the effects of conventional, cold nuclear matter on quarkonia yields, both of the nuclear modification of parton densities but also of possible absorption or final state effects. Note that quarkonia are suppressed also in $pA$ collisions, which constitutes one of the pieces of the small system puzzle. $eA$ collisions at the LHeC, with the possibility of varying the nuclear species, are a crucial part of the here sketched $hh/eh$ programme to clarify all these questions. 

Of special interest
is the physics of high transverse momentum particles and of jets, named {\it jet quenching}, usually employed in HI collisions as tools to analyse the QGP properties~\cite{Mehtar-Tani:2013pia} but of great interest in QCD and SM and for searches of BSM as well. The consideration of a complete detector, adding electromagnetic and hadronic calorimetry
and  muon detection to a superb inner tracking device,  opens numerous possibilities for studies of jet substructure, hadrochemistry and EM radiation within jets, heavy flavoured tagged jets, etc. 

It is to be noted that jet quenching is the only observation in HI that has not been found in small systems. $eA$ collisions at the LHeC offer the opportunity to study the influence of nuclear matter on jets~\cite{Li:2020rqj}, with abundant yields at high transverse momentum~\cite{AbelleiraFernandez:2012cc}, thus contributing to the understanding of the small system puzzle and of the physics of jets in HI collisions. A related subject is the use of high transverse momentum particles and jets to understand the initial stage of hadronic collisions~\cite{Apolinario:2017sob,Huss:2020dwe,Andres:2019eus}, an aspect that will benefit greatly from the possibility of varying the nuclear size of the colliding hadrons which provide larger centre-of-mass energies and luminosities~\cite{Citron:2018lsq}.

\subsubsection{Ultraperipheral Collisions}

\noindent Ultraperipheral collisions (UPC), in which one or both of the colliding hadrons act as sources of  large fluxes of quasi-real photons, are a hot topic at the LHC~\cite{Adam:2020mxg}. They offer the possibility of studying photo-production, being in that sense complementary to DIS in which the photon virtuality can be controlled and varied. They have been exploited until now through studies of exclusive vector meson production and dijets with the aim studying nuclear PDFs, and of dimuons and of two-particle correlations in the search of collective effects in systems smaller than $pp$\footnote{In this respect, the ATLAS Collaboration claims the observation of azimuthal asymmetries~\cite{Aad:2021yhy} in $\gamma$Pb collisions.}. UPC have also been used to study light-by-light scattering~\cite{Aad:2019ock}. All these possibilities can be further exploited in a new detector which besides tracking, calorimetry and muon detection, will provide photon, electron, proton and nucleus detection in the very backward and forward regions. $eA$ offers similar opportunities but in a much better controlled setup, with the additional possibility of further constraining the photon distribution inside the electron, see e.g.~\cite{Bertone:2019hks} and refs. therein. Further, the determination of the nuclear PDFs in inclusive processes in $eA$ would verify the numerous assumptions underlying their extraction in UPC.

\subsection{Detector concept for $eh$ and $hh$ collisions}
\label{subsec:ehandhh}

As described above in Section~\ref{Cocurrent_Operation}, the new accelerator optics is able to provide collisions for $eh$ and $hh$ configurations  in the same interaction point. As a consequence and if confirmed by further study, IP2 could indeed house one, common multipurpose detector serving for all of these, mostly related physics programs, of $ep,~pp,~eA,~pA$ and $AA$ interactions, with high precision and large acceptance, and the unique advantage for cross-calibration of performance and physics.

\begin{figure}[htb]
\begin{center}	
\includegraphics[width=0.49\textwidth]{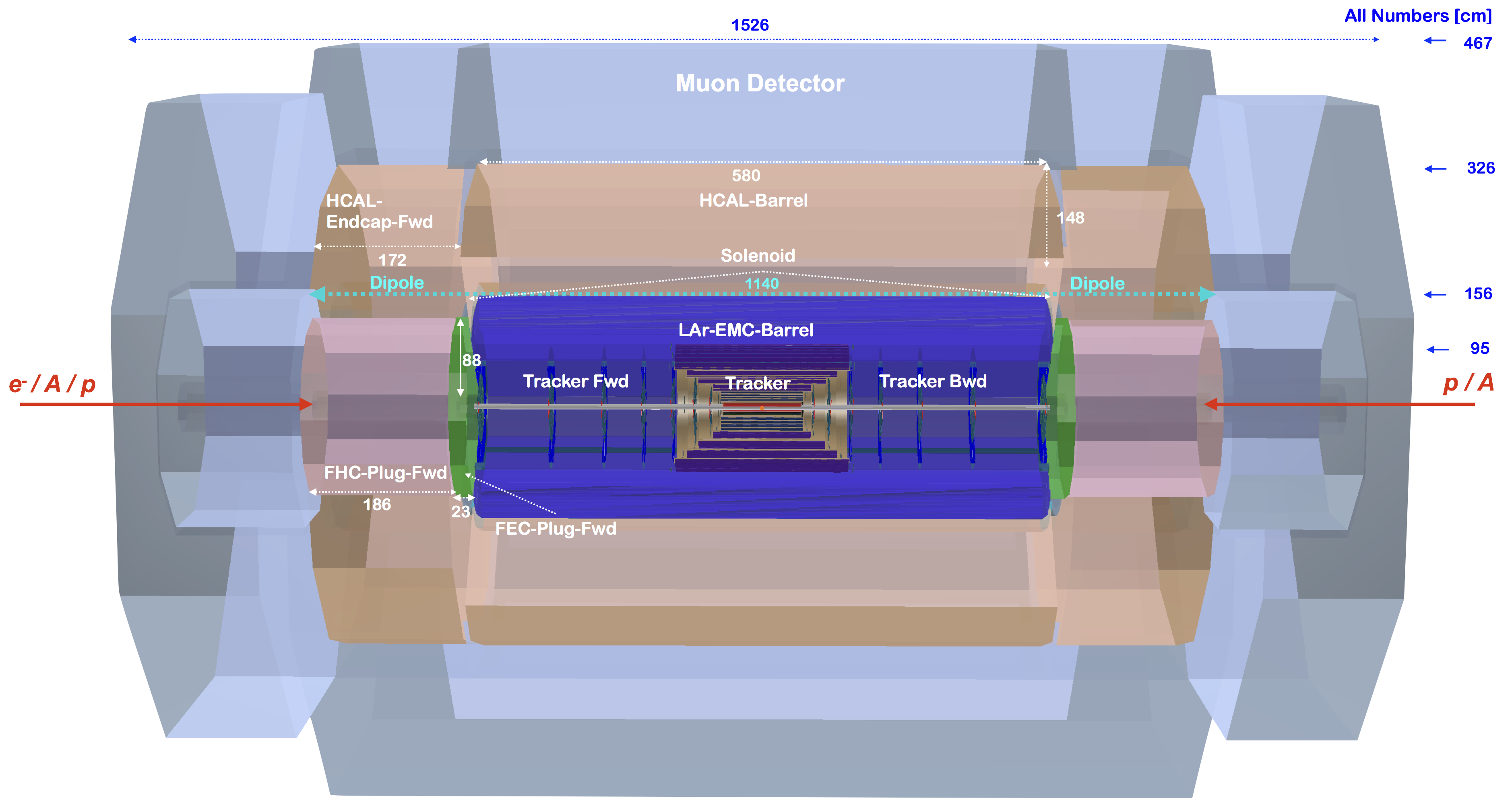}
\end{center}
\caption{Side view of a first design of the LHeC detector for both $eh$ and $hh$ collisions,
where the detector coverage of the backward (electron) direction is extended to match that for the forward (hadron) direction. The overall size in beam direction increased from 13.75\,m to 15.26\,m
while the diameter is unchanged as compared to the new $eh$ detector version presented
above.}
\label{fig:det:symLHeCdetector1}
\end{figure}

Exploiting the LHeC detector described above, 
a first design of a detector suitable for both $eh$ and $hh$ collisions
at IP2 is presented in Fig.~\ref{fig:det:symLHeCdetector1}. The detector is
symmetrised in the two beam directions, which is preferred
for $hh$ operation. The rapidity coverage of the electron side is
extended from $-4.6$ to $-5.2$ by adding two additional wheel tracking layers and
 a deeper and symmetric calorimetry. 
The solenoidal magnetic
field for normal $eh$ running is chosen to be  3\,T, while lower strengths
may also be considered.

The barrel, from inside out, comprises an inner Silicon tracker, surrounded by the electromagnetic (LAr)
calorimeter and a combined solenoid and dipole magnet section, at roughly 2\,m radius, 
followed by a hadronic calorimeter and muon system, both outside the 3\,T solenoidal field.
It extends to 4.67\,m radius, which fits into the L3 magnet support structure,
see~\cite{AbelleiraFernandez:2012cc} for a study on the installation of an LHeC detector
in the IP2 hall.

The LHeC central tracker provides momentum resolution
of about 1\% for $\eta < 3$ and $p_T$ below a few GeV or about 10\% for $\eta < 4.5$
and $p_T < 5$\,GeV. Detailed acceptance parameters are listed in Tab.\,1, where
now the backward detector parameters are to be replaced by those in forward direction.
The calorimetry also provides measurements for full rapidity coverage with fine segmentation, allowing to measure lepton, jets and energy flow to high precision as is 
illustrated in Section\,\ref{sec:calo}. The radial excursion of the LHeC barrel muon 
detector, as mentioned, is about $1.3$\,m.

Work on the LHeC detector concept continues, and more detailed simulations
would be a next step to study its performance.
The detail of the shape of the beam pipe for accommodating the synchrotron
radiation fan and hence the shape
of the circular-elliptical pixel layer are being worked out. A
further optimisation may also involve the size of strip layers for occupancy in $hh$
collisions and thickness of the calorimeters in backward direction. From this concept to
a technical design, work is ahead for a couple of years, linked also to the 
FCC feasibility study and the FCC-eh detector. Novel ideas, as are summarised in the
ECFA Detector R\&D Roadmap, such as on high resolution Silicon time measurements 
and integrated readout and wafer configurations are directions which will influence
the LHeC detector concept. 

\section{Summary}
\label{summary}

A detailed design has been presented and updated for 
the introduction of a programme 
of high energy electron-hadron scattering 
in a future phase of running of
the CERN Large Hadron Collider.
The design is based on collisions at interaction
point IP2, utilising one of the 
LHC hadron beams, and assumes
concurrent running with hadron-hadron
collider experiments. 

The electron beam as designed previously uses two
superconducting linear accelerators of 
length around 900\,m each, arranged in a 
racetrack configuration with three separate
return arcs, allowing acceleration 
to  an energy of 50\,GeV 
before bringing the beams into collision 
with the LHC hadrons. A key feature of the
electron accelerator 
design is energy recovery, for which  
plans for a lower energy, protoype ERL facility 
(PERLE) are well advanced.

The resulting LHeC experiment offers new
sensitivity to a broad and original programme of 
physics at the energy frontier also complementing
the existing LHC hadron-hadron 
experiments and their upgrades. 
Main physics topics are presented with a view on their 
requirements on the detector design.
Highlights of the 
high luminosity LHeC $ep$ programme 
picked out in this document include Higgs physics,
in a joint $ep$ and $pp$ coupling analysis,
studies of single top-quark production with
correspondingly high precision on the $Wtb$ vertex
and competitive 
sensitivity to physics beyond the standard model 
across a range of processes that benefit from
initial state leptons.

In terms of hadron structure, the LHeC allows
the extraction of the complete set of parton densities 
with unprecedented precision,
extending onto a new kinematic region, at low
Bjorken $x$ where new dynamics are expected
and to high $x$ where new physics may reside and factorisation
be tested independently of power and nuclear corrections.
In $eA$ mode, the LHeC offers unique sensitivity to nuclear
parton densities and exploits their enhanced
sensitivity to low $x$ effects over those of the
proton, as well as complementing the 
relativistic heavy ion collision programme at the 
LHC and RHIC by providing cold-matter baselines for the
understanding of quark-gluon plasma effects 
and contributing to a range of topics with hard probes
as well as soft physics and ultra-peripheral collisions. 

The ambitious physics programme is matched by a 
hermetic, compact, high performance
LHeC detector design based around a strong (3\,T) 
central solenoid for the precision measurement of
high transverse momentum charged particles. Inner
detectors based on depleted MAPS silicon sensors will
provide tracking and vertexing at the highest
possible precision with
a modest material budget. The pivotal importance
of scattered electron detection and measurement in
a DIS experiment is matched through 
electromagnetic calorimeter designs 
based on cold liquid argon with lead absorbers,
building on technologies used successfully in previous 
experiments. The need for a high quality hadron response,
from high transverse momentum
jets to the inclusive measurement of the
hadronic final state for kinematic reconstruction,
is met using a steel / scintillating tile solution. 
The importance of forward and very forward (and backward)
instrumentation is recognised by implementing 
detector components throughout the 
range $| \eta | < 5$ and by incorporating beamline 
instrumentation in the outgoing hadron and electron
directions into the interaction region design 
from the outset. 

Optimisations of the Interaction Region 
lead to improved performance parameters such as reduced synchrotron backgrounds
enabling the high luminosity goals to be pursued. New studies have been presented
on beam-beam effects and front-to-end tracking calculations
been made for different circumferences of the LHeC racetrack.

A first 
configuration has been developed of an IR which may alternately serve 
$eh$ and $hh$ collisions while the other experiments at the HL-LHC 
stay in $hh$ collision mode concurrently. The feasibility of realising 
a common $eh/hh$ IR implies that one may realistically consider 
an apparatus which would permit both 
electron-hadron and hadron-hadron collisions
to be registered.

The addition of
$ep$ and $eA$ experiment capabilities to the LHC accelerator
infrastructure, in combination with ongoing
$pp$ and $AA$ programmes, deepens the sensitivity
to new physics in the existing programme 
and adds TeV scale DIS physics in the quest for
finding new particles, dynamics or symmetries
beyond the Standard Model.
Whether as part of a multi-purpose apparatus or 
operating in standalone mode, the LHeC offers
fundamental, new perspectives on an energy frontier 
physics, detector and accelerator landscape in the 2030s which may indeed 
differ  from  today's perceptions.


\begin{acknowledgements}
This paper relies on a decade of work and collaboration with the hundreds of authors of the 2012~\cite{AbelleiraFernandez:2012cc} and 2020~\cite{Agostini:2020fmq} LHeC design papers. The work has been accompanied by the CERN Directorate and guided by an International Advisory Committee chaired by em. DG of CERN, Herwig Schopper.
Very fruitful discussions are acknowledged which some of us had with John Jowett, Luciano
Musa and further members of the ALICE Collaboration. Special thanks are due to the organisers
of the Off-Shell Conference, Kristin Lohwasser, Matthias Schott and colleagues, who encouraged and supported this paper to be written.

JGM gratefully acknowledges the hospitality of the CERN theory group.
We further acknowledge financial support by:
Xun\-ta de Galicia (Centro singular de investigaci\'on de Galicia accreditation 2019-2022); the "Mar\'{\i}a de Maeztu" Units of Excellence program MDM2016-0692 and the Spanish Research State Agency under project FPA2017-83814-P; Fun\-da\c cao para a Ci$\hat{\mathrm{e}}$ncia e a Tecnologia (Portugal) under project CERN/FISPAR/0024\-/2019; the South African Department of Science and Innovation and the National Research Foundation; the UK Science and Technology Facilities Council; U.S. DOE under contracts DE-AC05-06OR23177 and DE-SC0012704; European Union ERDF; the European Research Council under project
ERC-2018-ADG-835105 YoctoLHC; MSCA RISE 823947 "Heavy ion collisions: collectivity and precision in saturation physics"
(HIEIC); and European Union's Horizon 2020 research and innovation programme under
grant agreement No. 824093. 

\end{acknowledgements}



\end{document}